\documentclass[aps,preprint,onecolumn,superscriptaddress, shortbibliography, nobibnotes]{revtex4-2}

\usepackage{graphicx}

\usepackage{color}
\usepackage{float}
\usepackage{amsmath}

\begin{document}
	
	\title{Nanoscale modification of WS$_2$ trion emission by its local electromagnetic environment}
	
	\author{No\'emie Bonnet}
	\affiliation{Universit\'e Paris-Saclay, CNRS, Laboratoire de Physique des Solides, 91405, Orsay, France}
	
	\author{Hae Yeon Lee}
	\affiliation{Department of Materials Science and Engineering, Massachusetts Institute of Technology, 77 Massachusetts Ave, Cambridge, MA, 02141, USA}
	
	\author{Fuhui Shao}
	\affiliation{Universit\'e Paris-Saclay, CNRS, Laboratoire de Physique des Solides, 91405, Orsay, France}
	
	\author{Steffi Y. Woo}
	\affiliation{Universit\'e Paris-Saclay, CNRS, Laboratoire de Physique des Solides, 91405, Orsay, France}
	
	\author{Jean-Denis Blazit}
	\affiliation{Universit\'e Paris-Saclay, CNRS, Laboratoire de Physique des Solides, 91405, Orsay, France}
	
	\author{Kenji Watanabe}
	\affiliation{Research Center for Functional Materials, National Institute for Materials Science, 1-1 Namiki, Tsukuba 305-0044, Japan}
	
	\author{Takashi Taniguchi}
	\affiliation{International Center for Materials Nanoarchitectonics, National Institute for Materials Science,  1-1 Namiki, Tsukuba 305-0044, Japan}
	
	\author{Alberto Zobelli}
	\affiliation{Universit\'e Paris-Saclay, CNRS, Laboratoire de Physique des Solides, 91405, Orsay, France}
	
	\author{Odile St\'ephan}
	\affiliation{Universit\'e Paris-Saclay, CNRS, Laboratoire de Physique des Solides, 91405, Orsay, France}
	
	\author{Mathieu Kociak}
	\affiliation{Universit\'e Paris-Saclay, CNRS, Laboratoire de Physique des Solides, 91405, Orsay, France}
	
	\author{Silvija Gradecak-Garaj}
	\email{gradecak@nus.edu.sg}
	\affiliation{Department of Materials Science and Engineering, Massachusetts Institute of Technology, 77 Massachusetts Ave, Cambridge, MA, 02141, USA}
	
	\author{Luiz~H.~G.~Tizei}
	\email{luiz.galvao-tizei@universite-paris-saclay.fr}
	\affiliation{Universit\'e Paris-Saclay, CNRS, Laboratoire de Physique des Solides, 91405, Orsay, France}
	
	\date{\today}

	\begin{abstract}
		\textbf{Structural, electronic, and chemical nanoscale modifications of transition metal dichalcogenide monolayers alter their optical properties, including the generation of single photon emitters. A key missing element for complete control is a direct spatial correlation of optical response to nanoscale modifications, due to the large gap in spatial resolution between optical spectroscopy and nanometer resolved techniques, such as transmission electron microscopy or scanning tunneling microscopy. Here, we bridge this gap by obtaining nanometer resolved optical properties using electron spectroscopy, specifically electron energy loss spectroscopy (EELS) for absorption and cathodoluminescence (CL) for emission, which were directly correlated to chemical and structural information. In an h-BN/WS$_2$/h-BN heterostructure, we observe local modulation of the trion (X$^{-}$) emission due to tens of nanometer wide dielectric patches, while the exciton, X$_A$, does not follow the same modulation. Trion emission also increases in regions where charge accumulation occurs, close to the carbon film supporting the heterostructures. Finally, localized exciton emission (L) detection is not correlated to strain variations above 1 $\%$, suggesting point defects might be involved in their formations.}

	\end{abstract}

	\maketitle


	 Transition metal dichalcogenides (TMDs) of the form MX$_2$ (where M = W, Mo, and X = S, Se) with the 2H phase are semiconductors with indirect bandgap in bulk, and direct bandgap in monolayer  \cite{Mak2010}. Photoluminescence (PL) due to exciton decay is then brightest for monolayers. Their particular excitonic spin-valley physics, created by the lack of inversion symmetry, the strong spin-orbit coupling \cite{Xu2014}, and the reduced coulomb screening, have recently attracted great interest. Up to now, spectral changes in PL have not been linked to specific nanometer structural or chemical modifications in TMD monolayers, despite the observation of single photon emitters (SPE) in these materials. These SPE are of particular interest for their temporal stability, narrow spectral linewidths \cite{Tonndorf2015,Palacios2017, Darlington2020, Parto2020} indicating a low coupling to phonons, and possibility to create them selectively in space \cite{Palacios2017,Parto2020}, which places them as strong candidates for applications in quantum optics.

	 Near band-edge optical resonances of TMD monolayers are governed by excitonic transitions with Bohr radii in the nanometer range \cite{Molina2013}. The two lowest excitons, namely X$_A$ and X$_B$, occur at the $K$ point in reciprocal space and are split by spin-orbit coupling. Emission spectra of TMD monolayers, such as from PL spectroscopy, contain excitons (X$_A$) \cite{Arora2020}, trions that can be negatively- (X$^{-}$) \cite{Chernikov2015} or positively-charged (X$^{+}$) \cite{Paur2019}, and other lower energy lines, previously attributed to defects \cite{Arora2020} or potential changes due to strain \cite{Castellanos2013,Schmidt2016,Frisenda2017}. Some of these have been shown to be single photon emitters \cite{Tonndorf2015,Darlington2020}, which occur at energies below the X$^{-}$ emission, and are often referred as L peaks (for localized excitons) \cite{Jadczak2017, Koperski2017,Darlington2020}. Nanometer scale modulation of the dielectric environment of WSe$_2$ through a gated h-BN/graphene heterostructure creates moiré bands \cite{Xu2021}. These bands were attributed to local changes of the single particle bandgap, but not directly measured. Local measurements of the absorption of these heterostructures (where WSe$_2$ is buried in layers of h-BN and graphene) or the identification of the of nanoscale emitters in TMDs require at least an order of magnitude increase in the spatial resolution in measurements of the local structure and the chemistry to be coupled with optical measurements. Electron spectroscopies, such as electron energy loss spectroscopy (EELS), cathodoluminescence (CL) have the potential to address the obstacle of measuring the optical properties at deep sub-wavelength scales \cite{Polman2019}.

\begin{figure}
		\includegraphics[width=15cm]{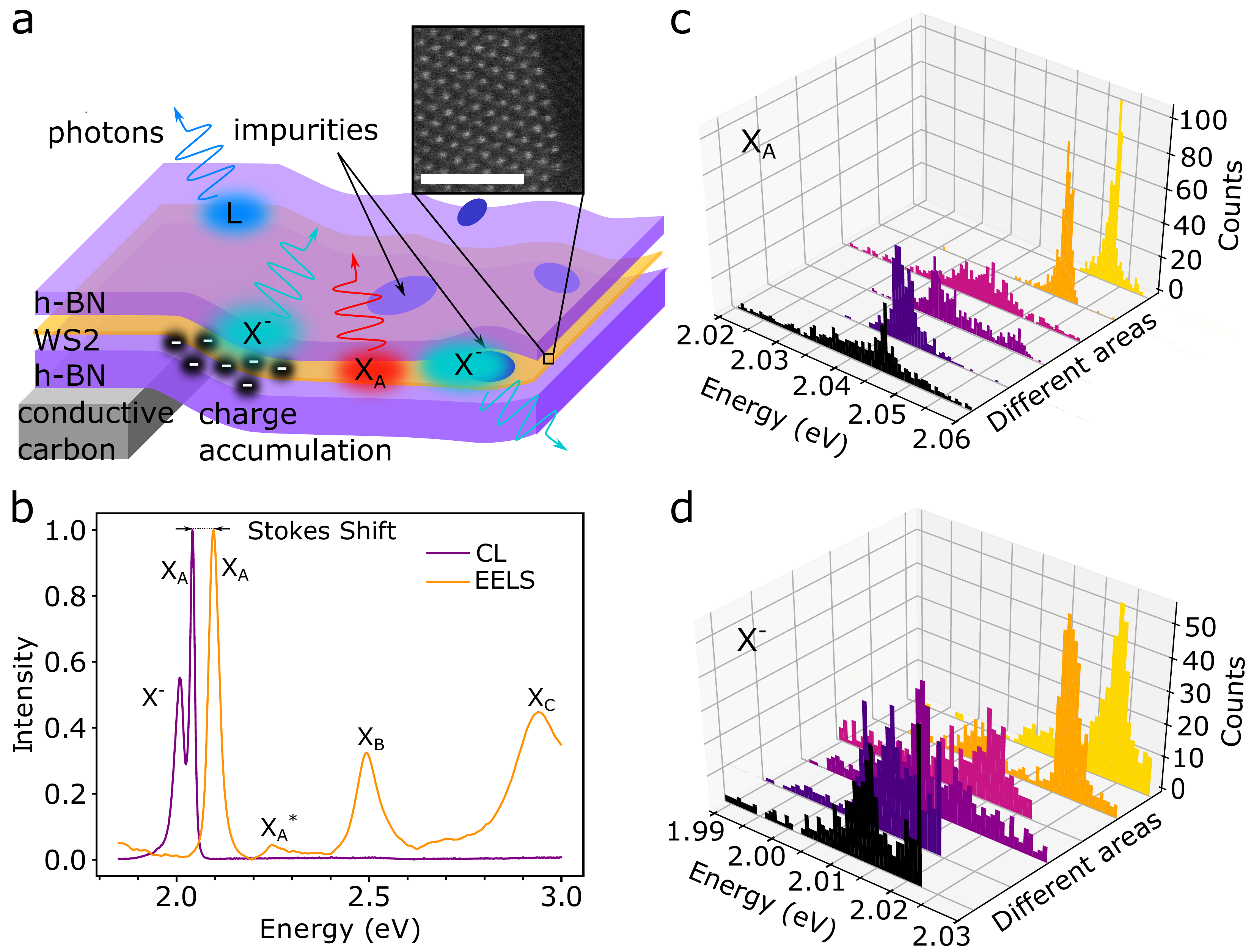}
		\caption{\textbf{Nanoscale optics of WS$_2$ monolayers:} \textbf{(a)} Sketch of the main results: a WS$_2$ monolayer (orange) encapsulated by two h-BN flakes (purple, 20 and 5 nm thickness) partially supported by a holey carbon film (gray) shows three emission lines: A excitons (X$_A$, red), trions (X$^{-}$, cyan) and localized emitters (L, blue). X$^{-}$ emission intensity increases close to the carbon support edges and around tens of nanometer residues patches throughout the WS$_2$ monolayer.  The inset shows an atomically resolved image of the WS$_2$ in the heterostructure (the scale bar is 2 nm). 
		\textbf{(b)} Typical electron energy loss spectra (EELS, orange) and cathodoluminescence (CL, purple). The X$_A$, X$_B$, X$_C$, and X$^{-}$ peaks are labeled. The extra absorption between X$_A$ and X$_B$ is attributed to the 2s excited state of X$_A$, marked X$_A^*$. The difference between X$_A$ emission and absorption maxima is marked as Stokes shift. The curves intensities are normalized to match the X$_A$ maxima. 
		\textbf{(c)} Histogram of the A exciton energy, measured by Gaussian fitting of the CL X$_A$ peak at each position contained in 6 different regions. \textbf{(d)} Histogram of the trion energy, measured by Gaussian fitting of the same CL data as (c). Each color represents a distinct region of between 1 and 2 µm$^2$ surface area containing hundreds of pixels, that would be only few pixels if measured by optical diffraction-limited methods.}
		\label{Figure_Experiment}
	\end{figure}

	Here, we used transmission electron microscopy techniques including highly monochromated EELS, CL, high angle annular dark field (HAADF), and diffraction imaging to correlate sub-10 nanometer optical absorption and luminescence spectral mapping to structural and chemical mapping. We show that the trion and other localized (L) light emission energy and intensity can vary on scales down to tens of nanometers in WS$_2$ monolayers while exciton emission intensity remains unchanged. In short, three effects are revealed: localized trion emission intensity increase at constant absorption rate due to either 1)  chemical changes in tens of nanometer patches or 2) to charge accumulation in a metal-insulator-semiconductor heterostructure combined with near-field emission enhancement; and 3) the presence of a bright emission, attributed to L, below the trion energy, highly localized in space. These effects, sketched in Fig. \ref{Figure_Experiment}a, are linked to local charge density variations and to near-field enhancement and not to the generally evoked local strain modification. In addition, the spatially-constant absorption, coupled to nanoscale-resolved CL, shows that the trion emission intensity increases due to a locally faster decay rate.

	X$_A$, X$^{-}$, and the localized emission possibly linked to defects have been observed with CL in TMD monolayers before \cite{Zheng2017,Nayak2019, Singh2020}, utilizing specific sample heterostructures, but with only hundreds of nanometer spatial resolution. More importantly, previous reports of EELS measurements on TMD monolayers did not reach spectral resolution comparable with optical absorption experiments to allow for facile interpretation \cite{Tizei2015, Habenicht2015, Nerl2017, Hong2020}. An alternative technique, scanning \textit{tunnelling} microscopy induced luminescence (STM-lum) allows the detection light emission from TMD monolayers \cite{PenaRoman2020,Schuler2020}. Despite the impressive atomic resolution achieved \cite{Schuler2020}, the strong influence of the STM tip on optical spectra hinders their use as an nanoscale equivalent of PL.

	The samples used here are made from WS$_2$ monolayers exfoliated from bulk material, and encapsulated in h-BN (5 and 25 nm thick on each side). The heterostructures created are then deposited on a conductive carbon film (ten of nanometer thick) with 2 $\mu$m-wide holes, itself supported on copper TEM grids (see the Methods section for details on sample preparation). The encapsulation offers high CL emission rate due to the increased interaction volume with the fast electron beam. The high charge-carriers density then produced enables CL detection \cite{Zheng2017,Nayak2019, Singh2020}. High purity and homogeneity in the h-BN layers are critical to get spectral line shapes comparable to those using pure optical means. Four samples were analyzed, with a typical surface area of 150 $\mu m^2$.

	Experiments were performed in a scanning transmission electron microscope (STEM), in which spatially resolved data are acquired by scanning a subnanometer electron beam, retrieving images (2D arrays with an intensity value in each pixel) and datacubes (2D arrays with a spectrum or a diffraction pattern at each pixel), see Fig. \ref{SI_experiment}. Energy filtered maps can be produced by cuts of these datacubes at different energies. Structural information was retrieved from atomic scale images (Fig. \ref{Figure_Experiment}a inset) and strain mapping through diffraction datacubes (Fig. \ref{SI_AtomicImage_Diffraction}). For more information, see the Methods section.

	Fig. \ref{Figure_Experiment}b presents the typical optical information that can be collected using CL and EELS on WS$_2$ monolayers with the samples kept at 150 K. CL and EELS spectral resolution used were 8 and 26 meV, respectively, for the measurements presented here.
	 
	 CL can be directly compared to off-resonance PL \cite{Mahfoud2013}, where the emission from X$_A$ and X$^{-}$ (X$^{+}$ occurs only in negatively gated WS$_2$ \cite{Paur2019}) is observed for the WS$_2$ monolayer. An overview of the energies measured for X$_A$ and X$^{-}$ CL emission is plotted as histograms in Fig. \ref{Figure_Experiment}c-d (more histograms corresponding to the localized L emission and exciton energies in EELS are shown in Fig. \ref{SI_histograms}), where the survey areas are of few $\mu$m each (represented by a different color). The measurements average around 2.049 eV and 2.011 eV (with full width at half maxima (FWHM) 13 meV and 19 meV, respectively), showing agreement between ensemble of CL measurements across a sample with macroscopic optical measurements (typical variations are of 20 to 30 meV for X$_A$ emission and absportion in regions above 10 $\mu$m$^2$ \cite{Jadczak2017, Arora2020, Kolesnichenko2020, Niehues2020}). Regions with broader ($\sim$15 meV, black histogram in Fig \ref{Figure_Experiment}c-d) and narrower ($\sim$5 meV, yellow histogram in the same figure) distributions can also be seen. In fact, of most interest in our study are the areas with spatial variations of the optical properties to unveil the origin of these variations.

	For atomically thin materials, EELS measures the imaginary part of the dielectric function \cite{Hambach2010, Kociak2017}, and it has been used for exciton mapping in TMDs \cite{Tizei2015, Habenicht2015, Nerl2017, Hong2020}. A comparison of the EELS and the optical absorbance ($1-R-T$, where $R$ and $T$ are the optical reflectivity and transmission \cite{Arora2020}) spectra of h-BN encapsulated WS$_2$ at 150 K shows a one-to-one match between features (see Fig. \ref{SI_EELS_Optics} and the Methods section for a discussion on EELS and optical absorption). The features seen in the EELS spectrum presented in Fig \ref{Figure_Experiment}b are exciton peaks. In addition to X$_A$ and X$_B$, two others are detected: X$_A^*$, the excited state of X$_A$, and X$_C$ which arises from strong absorption due to band nesting around the $Q$ point \cite{Carvalho2013}.	The energy positions of X$_A$ and X$_B$ measured in EELS are plotted as histograms in Fig. \ref{SI_histograms_holes} and Fig. \ref{SI_histograms}, showing the mean value of X$_A$ and X$_B$ to be 2.101 eV and 2.502 eV (with FWHM 24 meV and 25 meV, respectively).

\begin{figure}
		\includegraphics[width=14cm]{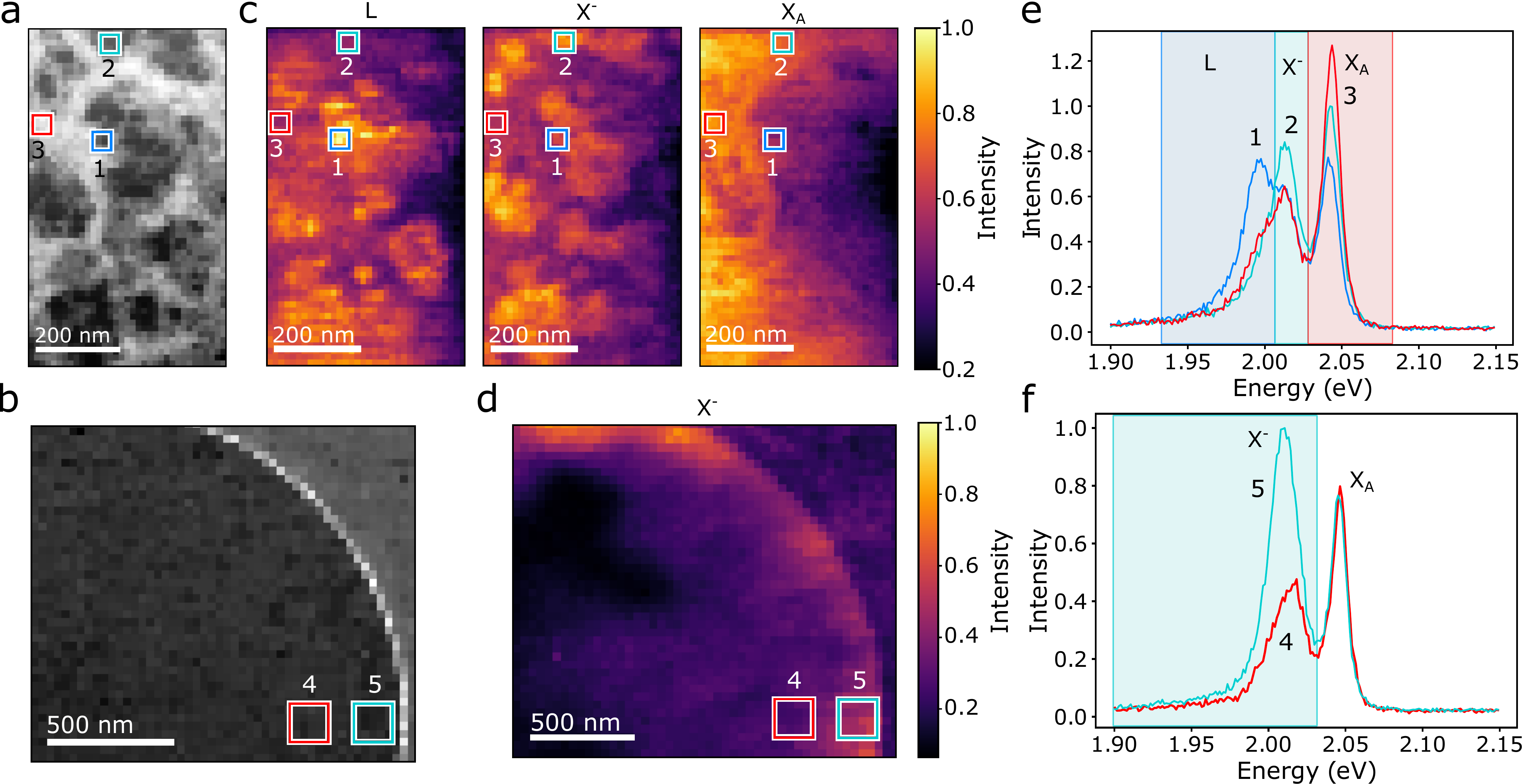}
		\caption{\textbf{Nanoscale emission intensity variations:} \textbf{(a)} HAADF image of the area measured in (c) and (e), \textbf{(b)} HAADF image of the suspended area measured in (d) and (f). \textbf{(c)} L emitters (left), X$^-$ (center) and X$_A$ (right) intensity maps, where small localized spots are seen for L and X$^-$. \textbf{(d)} X$^-$ intensity map showing trion enhancement next to the carbon support edge.\textbf{(e)} CL spectra corresponding to highlighted regions in (c). \textbf{(f)} CL spectra corresponding to highlighted regions in (d). The intensity in (e) and (f) was normalized by the maximum of the cyan spectrum to conserve the intensity changes of all peaks. The intensity in (c) and (d) was normalized by dividing by the maximum of each integrated datacube. The shaded regions in (e) and (f) mark where intensities were integrated for (c) and (d).}
		\label{Local_Trion_L1}
	\end{figure}

\begin{figure}
		\includegraphics[width=12cm]{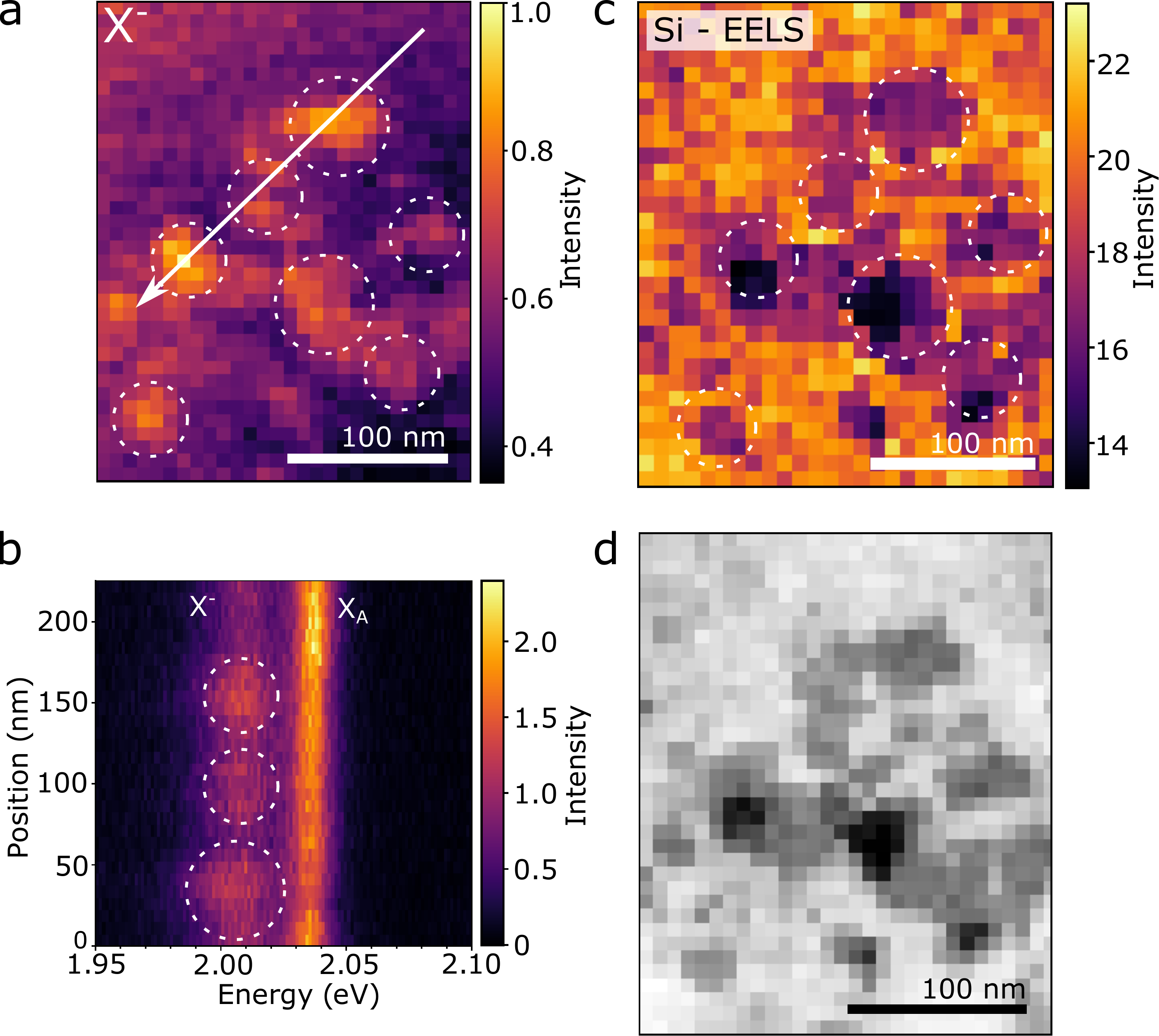}
		\caption{\textbf{Trion modification due to local surface patches: }  \textbf{(a)} X$^-$ intensity map. The intensity was normalized by the maximum of trion emission. \textbf{(b)} Spectral profile along the arrow in (a), where intensity modulation of X$^-$ occurs. \textbf{(c)} Residue content extracted using a blind source separation algorithm on the EELS datacube. The residue is present due to the monolayer transfer process. \textbf{(d)} HAADF image acquired in parallel with the EELS datacube image used to generate (c), with dark regions appearing to the lack of residue. X$^{-}$ maxima occur where the residue is not present.}
		\label{Trion_impurities}
	\end{figure}

\begin{figure}[h]
		\includegraphics[width=13cm]{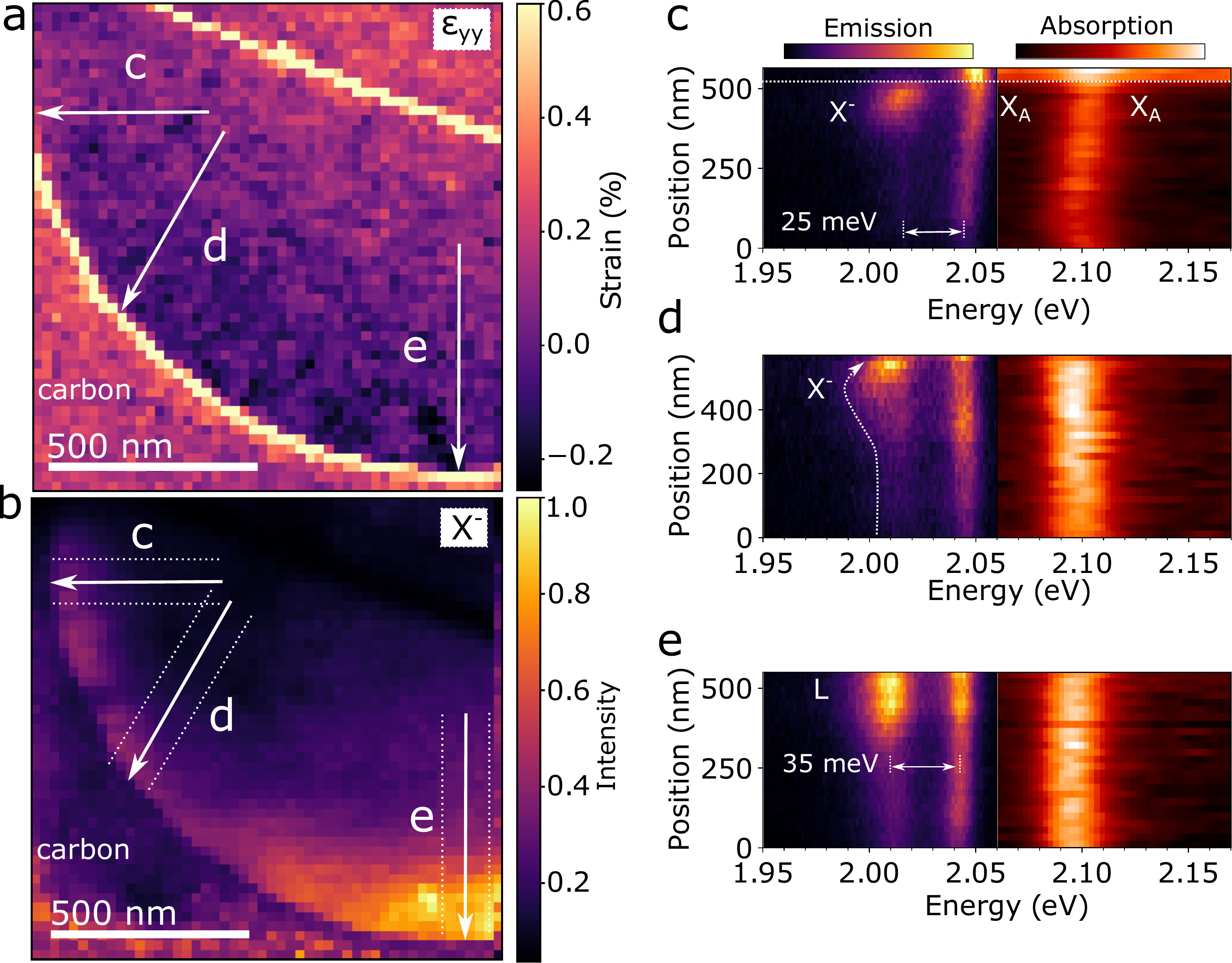}
		\caption{\textbf{Trion and L increase due to charge accumulation in a conductor-insulator-semiconductor interface:} \textbf{(a)} Strain map of $\epsilon_{yy}$ component. The brightest lines correspond to a fold (top line) and the edge of the carbon membrane (curved line). \textbf{(b)} X$^-$ intensity map, normalized by the maximum of trion emission. \textbf{(c-e)} CL (left) and EELS (right) spectra along each arrow marked in (a) and (b). In both (c) and (d) the trion peak redshifts then shifs back to its initial energy when approaching the carbon membrane (represented by a dotted line), as explained in the text. (e) A lower energy emission, marked by L 35 meV below the X$_A$ emission, which does not shift in energy is observed. It is attributed to localized excitons.}
		\label{Trion_interface}
	\end{figure}

	With this understanding of CL (EELS) as nanometer counterparts of PL (absorption), we can now describe our typical observations of deep sub-wavelength intensity variations, with examples shown in Fig. \ref{Local_Trion_L1}. X$^{-}$ and L peak intensities can vary locally in scales of tens of nanometers, while the X$_A$ peak intensity is relatively uniform. Fig. \ref{Local_Trion_L1}a and c show an example of this local intensity change for the X$^{-}$ and the L emission, which can occur for areas as small as 30x30 nm$^2$. Spectra averaged in such small regions (Fig. \ref{Local_Trion_L1}e) show varying peak heights. This behavior is not homogeneous across a single h-BN/WS$_2$/h-BN heterostructure or between different samples, and has been observed in regions of a few hundred nanometers across. A similar increase in X$^{-}$ intensity is observed close to edges of the carbon support (which appear as brighter regions in HAADF images, as in the upper right of Fig. \ref{Local_Trion_L1}b), which is in direct contact with the thicker (20 nm) h-BN layer (the sample details are described in the Methods section), but not with the WS$_2$ layer. Indeed, filtered emission maps at the trion energy (Fig. \ref{Local_Trion_L1}d) have stronger intensities close to the edge, as observed systematically in other holes of the same heterostructure  (see Fig. \ref{SI_TrionsOtherHoles}) and of different samples. Spectra close to the edge (cyan curve in Fig. \ref{Local_Trion_L1}f) show stronger trion emission in comparison to those in the suspended region (red curve in Fig. \ref{Local_Trion_L1}f). At first glance, these emission modifications could have the same origin. Yet, they occur at different scales Figs.  \ref{Trion_impurities} (tens of nanometers) and \ref{Trion_interface} (above a hundred nanometers) further entailing a detail analysis.
	
	In suspended regions where the X$^{-}$ intensity varies locally while X$_A$ intensity remains constant (Fig. \ref{Trion_impurities}a-b and Fig. \ref{SI_XA_intensity}), the typical spatial extension where enhanced trion emission is observed is of the order of tens of nanometers. As a function of position across different bright spots, the X$^{-}$ and X$_A$ intensities change independently (Fig. \ref{Trion_impurities}b), with no measurable energy shift. The typical size of these regions brings to mind the possibility of discrete light emitters, such as individual point defects, as observed in h-BN in the past \cite{Bourrellier2016} using CL. The trion formation and decay probabilities are known to depend on the local density of available free carriers, which can be modified not only by the presence of defects, but also by the local dielectric environment. HAADF images (Fig. \ref{Trion_impurities}d) of these regions show intensity variations, indicating the presence of extra matter either on the heterostructures' interfaces or surfaces. Core loss EELS shows that in addition to the expected chemical species (S, B, and, N), traces of impurities including Si, C, and O are also detected. Silicon, carbon, and oxygen impurities are expected residues from the sample preparation during the exfoliation of layers.
	
	 Blind source separation spectral analysis (see Methods, and Fig. \ref{SI_BSS}) shows that a component with Si, C, and O content is anti-correlated to the appearance of localized X$^{-}$ emission maxima: a map of this component is shown in Fig. \ref{Trion_impurities}c. The localized trion emission occurs in the areas which lack this residue-related component (marked by dash circles in Fig. \ref{Trion_impurities}). These same patches appear as minima in an HAADF image (Fig. \ref{Trion_impurities}d, which is proportional to the projected atomic number; see Fig. \ref{SI_BSS}b, Fig. \ref{Local_Trion_L1}a-b, and the Methods section). Their presence do not prevent the excitation transfer from the h-BN to the monolayer, indicating they are thin (as also suggested by EELS). h-BN/TMD stacks can be very clean \cite{Haigh2012}, but they contain some thin interface residue and bubbles. These additional surrounding dielectric patches change the local electromagnetic environment of the WS$_2$ monolayer. 
	
	 Around holes in the carbon support of the sample, the X$^{-}$ to X$_A$ ratio also increases (Fig. \ref{Trion_interface}a-b and Methods). On top of the carbon support the X$_A$ emission increases, with a drastic decrease of the X$^{-}$ to X$_A$ ratio. A series of emission spectra acquired starting from the suspended region up to the hole's edge along three different line profiles show the continuous increase in X$^{-}$ and X$_A$ emission. This evolution in emission occurs without observable modification of the absorption spectra (Fig. \ref{Trion_interface}c-e, right panels). The same behavior occurs in most of the analyzed holes in the carbon support (three other examples are shown in Fig. \ref{SI_TrionsOtherHoles}). A correlation between strain (Fig. \ref{Trion_interface}a shows the $\epsilon_{yy}$ component) and these modifications has not been detected (the bright lines occur due to structural changes including a fold and the carbon support edge).

	In addition to the X$^{-}$ intensity increase, its peak emission energy varies towards the edge of the support (white curved arrow in Fig. \ref{Trion_interface}d and Fig. \ref{SI_TrionsOtherHoles}): initially the peak redshifts by about 20 meV over a distance of 200 nm, and as its intensity increases, the redshift is followed by a final abrupt shift back to its initial energy, over a distance of 50 nm, and a larger intensity increase. The energy shifts lead to broader emission X$^{-}$ histograms compared the X$_A$ (orange curves in Fig. \ref{Figure_Experiment}c-d). Where this effect is observed, the X$_A$ emission and absorption energy do not follow the variations observed for the X$^{-}$. However, along with this characteristic shift of the trion, other energy shifts are observed, which the X$_A$ does follow (Fig. \ref{Trion_interface}c and Fig. \ref{SI_TrionsOtherHoles}). These will be disentangled in the discussion.

	Finally, close to the carbon support edge, one also observes localized emissions which match the lower-energy transitions referred to as L in literature \cite{Jadczak2017, Koperski2017, Arora2020,Paur2019}  (Fig. \ref{Trion_interface}, vertical profiles e in Fig. \ref{Trion_interface}a-b). This emission can be separated from that of the trion since their energy splittings to the X$_A$ are different: X$_A$ - X$^{-}$ is 35 meV  on average while X$_A$ - L is 45 meV. This particular energy splitting is systematically observed for this localized emission on the edge of the carbon support (observed on 21 measurements of 14 local emitters from two different samples). Its intensity is usually brighter than that of the trion, with an intensity ratio of $I_{L}/I_{X_A}$ = 3.4 on average, while it is of 1.3 for $I_{X^-}/I_{X_A}$. The width of the L emission is about the same as that of the X$^-$, respectively 31 and 33 meV, but larger than that of the X$_A$ is 17 meV, on average. The appearance of L emission  could not be directly linked to patterns in strain maps.
	

The ensemble of observations concerning the trion can be explained by making a hypothesis based on local changes of the free electron density and of the dielectric environment. Trion emission intensity can be controlled by gating of III-V \cite{Teran2005} and TMD \cite{Chernikov2015} semiconductors, which controls the density of free electrons; conversely, chemical doping can also modify this quantity \cite{Peimyoo2014}. Unintentional doping in MoS$_2$ has been shown to increase trion emission \cite{Neumann2018}, while a similar increase in WS$_2$ has been attributed to a larger concentration of defects \cite{Lin2018}. Substrate modification has also demonstrated an effect on the trion emission intensity in WS$_2$ monolayers \cite{Kobayashi2015}. In view of these reported observations, we attribute the localized trion emission increase described in Fig. \ref{Trion_impurities} to an augmentation of the local free electron density due to the absence of the surface contaminants. It has also been observed that strain (0.6 $\%$ and above) applied to WS$_2$ monolayers \cite{Harats2020} could induce trion intensity modification. Strain maps of regions around holes in the support do not show a correlation to the trion increase pattern (Fig. \ref{SI_strain_hole_local}).  Our strain measurements are not precise below 1$\%$ for the buried WS$_2$ monolayer (see Methods), so small deformations cannot be excluded. Finally, X$_A$ emission energy is not modified by the local dielectric patches. It is known that the optical bandgap of TMD monolayers is weakly influenced by the dielectric environment \cite{Xu2021}, as both the single particle bandgap and the exciton binding energy shift in tandem (to first order).

	At first sight, one could invoke a similar interpretation to the increase in the X$^{-}$/X$_A$ emission ratio close to the carbon support edge, that is, an effect of the local dielectric environment of the monolayer. However, given the sample geometry, the carbon support is not in contact with the monolayer, but separated by 20 nm of h-BN (the lower layer in the heterostructure), a thickness far larger than the extent of the exciton wavefunction outside the monolayer \cite{Molina2013}. More importantly, the amorphous carbon of the TEM grids is conductive, which is one of the reasons for their routine use in TEM, and it would quench light emission from the monolayers (TMD monolayer deposited directly on TEM grids do not emit light in CL experiments).
	
	It is exactly this conductive character of the carbon support that enters into play here. Our hypothesis is that the carbon support, h-BN, and WS$_2$ heterostructure forms a metal-insulator-semiconductor (MIS) capacitor, as the WS$_2$ is in contact with the carbon support away from measurement area. Therefore, free carriers at the center of the suspended region (distant from the carbon support) have different potential profiles than those in proximity to the carbon support. This changes the free carrier density at the center of the hole and around its edges, resulting in the different X$^{-}$/X$_A$ emission ratio. Rough estimates show that the capacitance created by a 20 nm h-BN (considering its bulk dielectric function), given the difference in workfunction between amorphous carbon and WS$_2$, can induce charge densities of the order of 1x$10^{13}$ cm$^{-2}$. A $3.2$x$10^{12}$ cm$^{-2}$ increase in the electron density in WS$_2$ (achieved by a 40 V gate voltage) has demonstrated an increase of the X$^{-}$ absorption intensity and a redshift of about 20 meV \cite{Chernikov2015}. This energy shift matches the magnitude of that observed in Fig. \ref{Trion_interface}c. In short, we interpret the redshift and higher emission rate as an increase in the trion population due to higher free electron concentration.

	A second effect is observed, in addition to the redshift of the trion energy over 200 nm when approaching the carbon support; a blueshift of the trion which is in fact a shift back to its initial energy far from the carbon membrane. Along with this shift, an intensity increase at distances below 50 nm from the edge of the membrane is clearly visible. This is at first  counterintuitive, as the former MIS capacitor explanation implies a continuous redshift with intensity increase. This second shift cannot be explained by a local change of the optical bandgap, since the energy of X$_A$ remains constant or changes marginally, both in absorption and emission (Fig. \ref{Trion_interface}d) across the whole range. A simple reduction of the charge density would explain the shift back to initial energy, but not the increase in emission. We attribute this second shift \textit{and} increase in intensity to a locally higher density of optical modes, which increases the decay rate due to the Purcell effect, where both the X$^{-}$ and X$_A$ are modified. A substantial increase in the emission intensity of molecules has been known to occur in close proximity to metallic structures \cite{Anger2006}.
	
	 In fact, such enhancement induces shorter exciton and trion lifetimes, leading directly to higher emission rates. It also explains the shift back in energy, which is related to the subsequent decrease of the trion population. We note that this is followed by an increase of X$_A$ emission, which is stronger on top of the carbon support (Fig. \ref{Trion_interface}c, above the white dashed line), where the trion emission is reduced (Fig. \ref{Trion_interface}d and \ref{SI_TrionsOtherHoles}). We interpret this as a consequence of the reduction of the X$_A$ lifetime, which increases its emission rate and decreases the trion formation probability. Here, we note that absorption intensity does not increase, ruling out a larger emission intensity simply due to a larger excitation rate. That is, at a constant excitation rate, the total number of exciton and trion formation is fixed, leading to a competition between their emission intensities.
	
	More specifically, the hypothesis of a trion blueshift due to strain is excluded because strain would also blueshift the X$_A$ emission and absorption energies, since it would change the WS$_2$ optical bandgap. Notably, in other regions X$_A$ emission and absorption are observed to change locally, as in the profiles in Fig. \ref{Trion_interface}c and Fig. \ref{SI_TrionsOtherHoles}. These energy shifts are followed by the X$^{-}$, but they do not preclude the general behavior demonstrated in Fig. \ref{Trion_interface}d. Strain mapping (Fig. \ref{SI_strain_hole_H1S1}) along the profiles in the regions where the X$_A$ energy (and X$^{-}$) varies does not allow one to attribute the energy profiles solely to strain. Shear and tensile strain and in-plane rotation are observed, but a one-to-one correspondence between these and the energy variations was not detected in Fig. \ref{SI_strain_hole_H1S1} and Fig. \ref{SI_strain_hole_H3S1}.

	Finally, we return to the L emission observation. Strain maps of the monolayer close to the carbon support edges show that it is strained (see Fig. \ref{SI_strain_hole_H3S1} and Fig. \ref{SI_strain_hole_H1S1}). The strain pattern is not simply that of a suspended membrane covering a circular hole, as one might initially expect. Indeed, regions close to the support edges show they can be under compression, including those where trion emission is increased. We interpret this as a result of the strain created during the heterostructure transfer.
	
	This complex strain profile brings to mind the multiple observations of single photon emitters in TMDs \cite{Tonndorf2015,Palacios2017,Darlington2020, Parto2020}, specifically WSe$_2$, which are currently attributed to the formation of localized excitonic states due to confinement.  Previous experiments in suspended layers \cite{Tonndorf2015} and layers deposited over nanopillars \cite{Palacios2017, Parto2020} show that strained layers lead to the formation of these single photon source. We do not observe a one-to-one correspondence of the appearance of L emission and strain maps. These emitters can be distinguished from trions based on their energy (they appear as distinct peaks in the binding energy histograms in Fig. \ref{SI_histograms_holes}) and have spatial localization below 100 nm, similarly to single photon emitters detected in h-BN using CL \cite{Bourrellier2016}. These and other localized emitters in TMDs warrant further exploration at the nanometer and atomic scales.
	
	In addition to allowing us to validate some hypothesis concerning energy shifts, the absorption and emission profiles shown in Fig. \ref{Trion_interface} give a local measure of the Stokes shift, the energy difference between emission and absorption of the same transition (X$_A$ here).  In molecular systems this energy difference occurs due to the interaction with phonons. In semiconductors, in addition to phonon interaction, other phenomena can intervene, such as doping, strain, and substrate-related effects. The Stokes shift measured for our samples is of the order of 40 meV. This is much larger than the smallest reported values for bare or h-BN encapsulated WS$_2$ monolayer \cite{Arora2020, Niehues2020}. We attribute this difference to sample quality (source of WS$_2$ or the heterostructure preparation), which further motivates future EELS and optical absorption experiments on the same objects.
	
	The results presented here demonstrate the existence of nanometer scale localized light emission in relatively structurally homogeneous TMD monolayers, which can be attributed to variations of the free electron density in the material caused by surface residue modifying the dielectric environment locally. Trion mapping on TMDs could also be used as local dielectric sensor, similar to the suggestion by Xu et. al \cite{Xu2021} based on optical reflectivity. From another perspective, the creation of nanoscale emitters indicate that  dense arrays could be engineered by manipulation of the surface, such as by way of patterning. Finally, a lack of correlation between L emitters and strain above 1$\%$ indicates that strain alone is not sufficient for their generation. Possibly, point defects are necessary to generate them, as suggested by the detection of single photon emitters in h-BN encapsulated WSe$_2$ placed on dielectric pillars only after 100 keV electron-irradiation \cite{Parto2020}. As such, nanoscale electron microscopy and spectroscopy can offer a way to generate and characterize atomic-scale defects, and to monitor the change in optical response in real-time towards better understanding of nanoscale emitters in TMDs.

	\bibliography{Bib_WS2_NanoscaleTrion.bib}

\begin{thebibliography}{54}%
\makeatletter
\providecommand \@ifxundefined [1]{%
 \@ifx{#1\undefined}
}%
\providecommand \@ifnum [1]{%
 \ifnum #1\expandafter \@firstoftwo
 \else \expandafter \@secondoftwo
 \fi
}%
\providecommand \@ifx [1]{%
 \ifx #1\expandafter \@firstoftwo
 \else \expandafter \@secondoftwo
 \fi
}%
\providecommand \natexlab [1]{#1}%
\providecommand \enquote  [1]{``#1''}%
\providecommand \bibnamefont  [1]{#1}%
\providecommand \bibfnamefont [1]{#1}%
\providecommand \citenamefont [1]{#1}%
\providecommand \href@noop [0]{\@secondoftwo}%
\providecommand \href [0]{\begingroup \@sanitize@url \@href}%
\providecommand \@href[1]{\@@startlink{#1}\@@href}%
\providecommand \@@href[1]{\endgroup#1\@@endlink}%
\providecommand \@sanitize@url [0]{\catcode `\\12\catcode `\$12\catcode
  `\&12\catcode `\#12\catcode `\^12\catcode `\_12\catcode `\%12\relax}%
\providecommand \@@startlink[1]{}%
\providecommand \@@endlink[0]{}%
\providecommand \url  [0]{\begingroup\@sanitize@url \@url }%
\providecommand \@url [1]{\endgroup\@href {#1}{\urlprefix }}%
\providecommand \urlprefix  [0]{URL }%
\providecommand \Eprint [0]{\href }%
\providecommand \doibase [0]{https://doi.org/}%
\providecommand \selectlanguage [0]{\@gobble}%
\providecommand \bibinfo  [0]{\@secondoftwo}%
\providecommand \bibfield  [0]{\@secondoftwo}%
\providecommand \translation [1]{[#1]}%
\providecommand \BibitemOpen [0]{}%
\providecommand \bibitemStop [0]{}%
\providecommand \bibitemNoStop [0]{.\EOS\space}%
\providecommand \EOS [0]{\spacefactor3000\relax}%
\providecommand \BibitemShut  [1]{\csname bibitem#1\endcsname}%
\let\auto@bib@innerbib\@empty
\bibitem [{\citenamefont {Mak}\ \emph {et~al.}(2010)\citenamefont {Mak},
  \citenamefont {Lee}, \citenamefont {Hone}, \citenamefont {Shan},\ and\
  \citenamefont {Heinz}}]{Mak2010}%
  \BibitemOpen
  \bibfield  {author} {\bibinfo {author} {\bibfnamefont {K.~F.}\ \bibnamefont
  {Mak}}, \bibinfo {author} {\bibfnamefont {C.}~\bibnamefont {Lee}}, \bibinfo
  {author} {\bibfnamefont {J.}~\bibnamefont {Hone}}, \bibinfo {author}
  {\bibfnamefont {J.}~\bibnamefont {Shan}},\ and\ \bibinfo {author}
  {\bibfnamefont {T.~F.}\ \bibnamefont {Heinz}},\ }\bibfield  {title} {\bibinfo
  {title} {Atomically thin {MoS}$_{2}$: A new direct-gap semiconductor},\
  }\href {https://doi.org/10.1103/PhysRevLett.105.136805} {\bibfield  {journal}
  {\bibinfo  {journal} {Phys. Rev. Lett.}\ }\textbf {\bibinfo {volume} {105}},\
  \bibinfo {pages} {136805} (\bibinfo {year} {2010})}\BibitemShut {NoStop}%
\bibitem [{\citenamefont {Xu}\ \emph {et~al.}(2014)\citenamefont {Xu},
  \citenamefont {Yao}, \citenamefont {Xiao},\ and\ \citenamefont
  {Heinz}}]{Xu2014}%
  \BibitemOpen
  \bibfield  {author} {\bibinfo {author} {\bibfnamefont {X.}~\bibnamefont
  {Xu}}, \bibinfo {author} {\bibfnamefont {W.}~\bibnamefont {Yao}}, \bibinfo
  {author} {\bibfnamefont {D.}~\bibnamefont {Xiao}},\ and\ \bibinfo {author}
  {\bibfnamefont {T.~F.}\ \bibnamefont {Heinz}},\ }\bibfield  {title} {\bibinfo
  {title} {Spin and pseudospins in layered transition metal dichalcogenides},\
  }\href@noop {} {\bibfield  {journal} {\bibinfo  {journal} {Nat. Phys.}\
  }\textbf {\bibinfo {volume} {10}},\ \bibinfo {pages} {343} (\bibinfo {year}
  {2014})}\BibitemShut {NoStop}%
\bibitem [{\citenamefont {Tonndorf}\ \emph {et~al.}(2015)\citenamefont
  {Tonndorf}, \citenamefont {Schmidt}, \citenamefont {Schneider}, \citenamefont
  {Kern}, \citenamefont {Buscema}, \citenamefont {Steele}, \citenamefont
  {Castellanos-Gomez}, \citenamefont {van~der Zant}, \citenamefont
  {de~Vasconcellos},\ and\ \citenamefont {Bratschitsch}}]{Tonndorf2015}%
  \BibitemOpen
  \bibfield  {author} {\bibinfo {author} {\bibfnamefont {P.}~\bibnamefont
  {Tonndorf}}, \bibinfo {author} {\bibfnamefont {R.}~\bibnamefont {Schmidt}},
  \bibinfo {author} {\bibfnamefont {R.}~\bibnamefont {Schneider}}, \bibinfo
  {author} {\bibfnamefont {J.}~\bibnamefont {Kern}}, \bibinfo {author}
  {\bibfnamefont {M.}~\bibnamefont {Buscema}}, \bibinfo {author} {\bibfnamefont
  {G.~A.}\ \bibnamefont {Steele}}, \bibinfo {author} {\bibfnamefont
  {A.}~\bibnamefont {Castellanos-Gomez}}, \bibinfo {author} {\bibfnamefont
  {H.~S.}\ \bibnamefont {van~der Zant}}, \bibinfo {author} {\bibfnamefont
  {S.~M.}\ \bibnamefont {de~Vasconcellos}},\ and\ \bibinfo {author}
  {\bibfnamefont {R.}~\bibnamefont {Bratschitsch}},\ }\bibfield  {title}
  {\bibinfo {title} {Single-photon emission from localized excitons in an
  atomically thin semiconductor},\ }\href@noop {} {\bibfield  {journal}
  {\bibinfo  {journal} {Optica}\ }\textbf {\bibinfo {volume} {2}},\ \bibinfo
  {pages} {347} (\bibinfo {year} {2015})}\BibitemShut {NoStop}%
\bibitem [{\citenamefont {Palacios-Berraquero}\ \emph
  {et~al.}(2017)\citenamefont {Palacios-Berraquero}, \citenamefont {Kara},
  \citenamefont {Montblanch}, \citenamefont {Barbone}, \citenamefont
  {Latawiec}, \citenamefont {Yoon}, \citenamefont {Ott}, \citenamefont
  {Loncar}, \citenamefont {Ferrari},\ and\ \citenamefont
  {Atat{\"u}re}}]{Palacios2017}%
  \BibitemOpen
  \bibfield  {author} {\bibinfo {author} {\bibfnamefont {C.}~\bibnamefont
  {Palacios-Berraquero}}, \bibinfo {author} {\bibfnamefont {D.~M.}\
  \bibnamefont {Kara}}, \bibinfo {author} {\bibfnamefont {A.~R.-P.}\
  \bibnamefont {Montblanch}}, \bibinfo {author} {\bibfnamefont
  {M.}~\bibnamefont {Barbone}}, \bibinfo {author} {\bibfnamefont
  {P.}~\bibnamefont {Latawiec}}, \bibinfo {author} {\bibfnamefont
  {D.}~\bibnamefont {Yoon}}, \bibinfo {author} {\bibfnamefont {A.~K.}\
  \bibnamefont {Ott}}, \bibinfo {author} {\bibfnamefont {M.}~\bibnamefont
  {Loncar}}, \bibinfo {author} {\bibfnamefont {A.~C.}\ \bibnamefont
  {Ferrari}},\ and\ \bibinfo {author} {\bibfnamefont {M.}~\bibnamefont
  {Atat{\"u}re}},\ }\bibfield  {title} {\bibinfo {title} {Large-scale
  quantum-emitter arrays in atomically thin semiconductors},\ }\href@noop {}
  {\bibfield  {journal} {\bibinfo  {journal} {Nature Comm.}\ }\textbf {\bibinfo
  {volume} {8}},\ \bibinfo {pages} {1} (\bibinfo {year} {2017})}\BibitemShut
  {NoStop}%
\bibitem [{\citenamefont {Darlington}\ \emph {et~al.}(2020)\citenamefont
  {Darlington}, \citenamefont {Carmesin}, \citenamefont {Florian},
  \citenamefont {Yanev}, \citenamefont {Ajayi}, \citenamefont {Ardelean},
  \citenamefont {Rhodes}, \citenamefont {Ghiotto}, \citenamefont {Krayev},
  \citenamefont {Watanabe} \emph {et~al.}}]{Darlington2020}%
  \BibitemOpen
  \bibfield  {author} {\bibinfo {author} {\bibfnamefont {T.~P.}\ \bibnamefont
  {Darlington}}, \bibinfo {author} {\bibfnamefont {C.}~\bibnamefont
  {Carmesin}}, \bibinfo {author} {\bibfnamefont {M.}~\bibnamefont {Florian}},
  \bibinfo {author} {\bibfnamefont {E.}~\bibnamefont {Yanev}}, \bibinfo
  {author} {\bibfnamefont {O.}~\bibnamefont {Ajayi}}, \bibinfo {author}
  {\bibfnamefont {J.}~\bibnamefont {Ardelean}}, \bibinfo {author}
  {\bibfnamefont {D.~A.}\ \bibnamefont {Rhodes}}, \bibinfo {author}
  {\bibfnamefont {A.}~\bibnamefont {Ghiotto}}, \bibinfo {author} {\bibfnamefont
  {A.}~\bibnamefont {Krayev}}, \bibinfo {author} {\bibfnamefont
  {K.}~\bibnamefont {Watanabe}}, \emph {et~al.},\ }\bibfield  {title} {\bibinfo
  {title} {Imaging strain-localized exciton states in nanoscale bubbles in
  monolayer {WSe}$_2$ at room temperature},\ }\href@noop {} {\bibfield
  {journal} {\bibinfo  {journal} {Nat. Nano}\ }\textbf {\bibinfo {volume}
  {15}},\ \bibinfo {pages} {854} (\bibinfo {year} {2020})}\BibitemShut
  {NoStop}%
\bibitem [{\citenamefont {Parto}\ \emph {et~al.}(2020)\citenamefont {Parto},
  \citenamefont {Banerjee},\ and\ \citenamefont {Moody}}]{Parto2020}%
  \BibitemOpen
  \bibfield  {author} {\bibinfo {author} {\bibfnamefont {K.}~\bibnamefont
  {Parto}}, \bibinfo {author} {\bibfnamefont {K.}~\bibnamefont {Banerjee}},\
  and\ \bibinfo {author} {\bibfnamefont {G.}~\bibnamefont {Moody}},\
  }\href@noop {} {\bibinfo {title} {Irradiation of nanostrained monolayer
  {WSe}$_2$ for site-controlled single-photon emission up to 150 k}} (\bibinfo
  {year} {2020}),\ \Eprint {https://arxiv.org/abs/2009.07315} {arXiv:2009.07315
  [physics.app-ph]} \BibitemShut {NoStop}%
\bibitem [{\citenamefont {Molina-S\'anchez}\ \emph {et~al.}(2013)\citenamefont
  {Molina-S\'anchez}, \citenamefont {Sangalli}, \citenamefont {Hummer},
  \citenamefont {Marini},\ and\ \citenamefont {Wirtz}}]{Molina2013}%
  \BibitemOpen
  \bibfield  {author} {\bibinfo {author} {\bibfnamefont {A.}~\bibnamefont
  {Molina-S\'anchez}}, \bibinfo {author} {\bibfnamefont {D.}~\bibnamefont
  {Sangalli}}, \bibinfo {author} {\bibfnamefont {K.}~\bibnamefont {Hummer}},
  \bibinfo {author} {\bibfnamefont {A.}~\bibnamefont {Marini}},\ and\ \bibinfo
  {author} {\bibfnamefont {L.}~\bibnamefont {Wirtz}},\ }\bibfield  {title}
  {\bibinfo {title} {Effect of spin-orbit interaction on the optical spectra of
  single-layer, double-layer, and bulk {M}o{S}$_{2}$},\ }\href@noop {}
  {\bibfield  {journal} {\bibinfo  {journal} {Phys. Rev. B}\ }\textbf {\bibinfo
  {volume} {88}},\ \bibinfo {pages} {045412} (\bibinfo {year}
  {2013})}\BibitemShut {NoStop}%
\bibitem [{\citenamefont {Arora}\ \emph {et~al.}(2020)\citenamefont {Arora},
  \citenamefont {Wessling}, \citenamefont {Deilmann}, \citenamefont
  {Reichenauer}, \citenamefont {Steeger}, \citenamefont {Kossacki},
  \citenamefont {Potemski}, \citenamefont {de~Vasconcellos}, \citenamefont
  {Rohlfing},\ and\ \citenamefont {Bratschitsch}}]{Arora2020}%
  \BibitemOpen
  \bibfield  {author} {\bibinfo {author} {\bibfnamefont {A.}~\bibnamefont
  {Arora}}, \bibinfo {author} {\bibfnamefont {N.~K.}\ \bibnamefont {Wessling}},
  \bibinfo {author} {\bibfnamefont {T.}~\bibnamefont {Deilmann}}, \bibinfo
  {author} {\bibfnamefont {T.}~\bibnamefont {Reichenauer}}, \bibinfo {author}
  {\bibfnamefont {P.}~\bibnamefont {Steeger}}, \bibinfo {author} {\bibfnamefont
  {P.}~\bibnamefont {Kossacki}}, \bibinfo {author} {\bibfnamefont
  {M.}~\bibnamefont {Potemski}}, \bibinfo {author} {\bibfnamefont {S.~M.}\
  \bibnamefont {de~Vasconcellos}}, \bibinfo {author} {\bibfnamefont
  {M.}~\bibnamefont {Rohlfing}},\ and\ \bibinfo {author} {\bibfnamefont
  {R.}~\bibnamefont {Bratschitsch}},\ }\bibfield  {title} {\bibinfo {title}
  {Dark trions govern the temperature-dependent optical absorption and emission
  of doped atomically thin semiconductors},\ }\href@noop {} {\bibfield
  {journal} {\bibinfo  {journal} {Phys. Rev. B}\ }\textbf {\bibinfo {volume}
  {101}},\ \bibinfo {pages} {241413} (\bibinfo {year} {2020})}\BibitemShut
  {NoStop}%
\bibitem [{\citenamefont {Chernikov}\ \emph {et~al.}(2015)\citenamefont
  {Chernikov}, \citenamefont {van~der Zande}, \citenamefont {Hill},
  \citenamefont {Rigosi}, \citenamefont {Velauthapillai}, \citenamefont
  {Hone},\ and\ \citenamefont {Heinz}}]{Chernikov2015}%
  \BibitemOpen
  \bibfield  {author} {\bibinfo {author} {\bibfnamefont {A.}~\bibnamefont
  {Chernikov}}, \bibinfo {author} {\bibfnamefont {A.~M.}\ \bibnamefont {van~der
  Zande}}, \bibinfo {author} {\bibfnamefont {H.~M.}\ \bibnamefont {Hill}},
  \bibinfo {author} {\bibfnamefont {A.~F.}\ \bibnamefont {Rigosi}}, \bibinfo
  {author} {\bibfnamefont {A.}~\bibnamefont {Velauthapillai}}, \bibinfo
  {author} {\bibfnamefont {J.}~\bibnamefont {Hone}},\ and\ \bibinfo {author}
  {\bibfnamefont {T.~F.}\ \bibnamefont {Heinz}},\ }\bibfield  {title} {\bibinfo
  {title} {Electrical tuning of exciton binding energies in monolayer
  {WS}$_{2}$},\ }\href {https://doi.org/10.1103/PhysRevLett.115.126802}
  {\bibfield  {journal} {\bibinfo  {journal} {Phys. Rev. Lett.}\ }\textbf
  {\bibinfo {volume} {115}},\ \bibinfo {pages} {126802} (\bibinfo {year}
  {2015})}\BibitemShut {NoStop}%
\bibitem [{\citenamefont {Paur}\ \emph {et~al.}(2019)\citenamefont {Paur},
  \citenamefont {Molina-Mendoza}, \citenamefont {Bratschitsch}, \citenamefont
  {Watanabe}, \citenamefont {Taniguchi},\ and\ \citenamefont
  {Mueller}}]{Paur2019}%
  \BibitemOpen
  \bibfield  {author} {\bibinfo {author} {\bibfnamefont {M.}~\bibnamefont
  {Paur}}, \bibinfo {author} {\bibfnamefont {A.~J.}\ \bibnamefont
  {Molina-Mendoza}}, \bibinfo {author} {\bibfnamefont {R.}~\bibnamefont
  {Bratschitsch}}, \bibinfo {author} {\bibfnamefont {K.}~\bibnamefont
  {Watanabe}}, \bibinfo {author} {\bibfnamefont {T.}~\bibnamefont
  {Taniguchi}},\ and\ \bibinfo {author} {\bibfnamefont {T.}~\bibnamefont
  {Mueller}},\ }\bibfield  {title} {\bibinfo {title} {Electroluminescence from
  multi-particle exciton complexes in transition metal dichalcogenide
  semiconductors},\ }\href@noop {} {\bibfield  {journal} {\bibinfo  {journal}
  {Nat. Comm.}\ }\textbf {\bibinfo {volume} {10}},\ \bibinfo {pages} {1}
  (\bibinfo {year} {2019})}\BibitemShut {NoStop}%
\bibitem [{\citenamefont {Castellanos-Gomez}\ \emph {et~al.}(2013)\citenamefont
  {Castellanos-Gomez}, \citenamefont {Rold{\'a}n}, \citenamefont {Cappelluti},
  \citenamefont {Buscema}, \citenamefont {Guinea}, \citenamefont {van~der
  Zant},\ and\ \citenamefont {Steele}}]{Castellanos2013}%
  \BibitemOpen
  \bibfield  {author} {\bibinfo {author} {\bibfnamefont {A.}~\bibnamefont
  {Castellanos-Gomez}}, \bibinfo {author} {\bibfnamefont {R.}~\bibnamefont
  {Rold{\'a}n}}, \bibinfo {author} {\bibfnamefont {E.}~\bibnamefont
  {Cappelluti}}, \bibinfo {author} {\bibfnamefont {M.}~\bibnamefont {Buscema}},
  \bibinfo {author} {\bibfnamefont {F.}~\bibnamefont {Guinea}}, \bibinfo
  {author} {\bibfnamefont {H.~S.}\ \bibnamefont {van~der Zant}},\ and\ \bibinfo
  {author} {\bibfnamefont {G.~A.}\ \bibnamefont {Steele}},\ }\bibfield  {title}
  {\bibinfo {title} {Local strain engineering in atomically thin {MoS}$_2$},\
  }\href@noop {} {\bibfield  {journal} {\bibinfo  {journal} {Nano Lett.}\
  }\textbf {\bibinfo {volume} {13}},\ \bibinfo {pages} {5361} (\bibinfo {year}
  {2013})}\BibitemShut {NoStop}%
\bibitem [{\citenamefont {Schmidt}\ \emph {et~al.}(2016)\citenamefont
  {Schmidt}, \citenamefont {Niehues}, \citenamefont {Schneider}, \citenamefont
  {Dr{\"u}ppel}, \citenamefont {Deilmann}, \citenamefont {Rohlfing},
  \citenamefont {De~Vasconcellos}, \citenamefont {Castellanos-Gomez},\ and\
  \citenamefont {Bratschitsch}}]{Schmidt2016}%
  \BibitemOpen
  \bibfield  {author} {\bibinfo {author} {\bibfnamefont {R.}~\bibnamefont
  {Schmidt}}, \bibinfo {author} {\bibfnamefont {I.}~\bibnamefont {Niehues}},
  \bibinfo {author} {\bibfnamefont {R.}~\bibnamefont {Schneider}}, \bibinfo
  {author} {\bibfnamefont {M.}~\bibnamefont {Dr{\"u}ppel}}, \bibinfo {author}
  {\bibfnamefont {T.}~\bibnamefont {Deilmann}}, \bibinfo {author}
  {\bibfnamefont {M.}~\bibnamefont {Rohlfing}}, \bibinfo {author}
  {\bibfnamefont {S.~M.}\ \bibnamefont {De~Vasconcellos}}, \bibinfo {author}
  {\bibfnamefont {A.}~\bibnamefont {Castellanos-Gomez}},\ and\ \bibinfo
  {author} {\bibfnamefont {R.}~\bibnamefont {Bratschitsch}},\ }\bibfield
  {title} {\bibinfo {title} {Reversible uniaxial strain tuning in atomically
  thin {WSe}$_2$},\ }\href@noop {} {\bibfield  {journal} {\bibinfo  {journal}
  {2D Materials}\ }\textbf {\bibinfo {volume} {3}},\ \bibinfo {pages} {021011}
  (\bibinfo {year} {2016})}\BibitemShut {NoStop}%
\bibitem [{\citenamefont {Frisenda}\ \emph {et~al.}(2017)\citenamefont
  {Frisenda}, \citenamefont {Dr{\"u}ppel}, \citenamefont {Schmidt},
  \citenamefont {de~Vasconcellos}, \citenamefont {de~Lara}, \citenamefont
  {Bratschitsch}, \citenamefont {Rohlfing},\ and\ \citenamefont
  {Castellanos-Gomez}}]{Frisenda2017}%
  \BibitemOpen
  \bibfield  {author} {\bibinfo {author} {\bibfnamefont {R.}~\bibnamefont
  {Frisenda}}, \bibinfo {author} {\bibfnamefont {M.}~\bibnamefont
  {Dr{\"u}ppel}}, \bibinfo {author} {\bibfnamefont {R.}~\bibnamefont
  {Schmidt}}, \bibinfo {author} {\bibfnamefont {S.~M.}\ \bibnamefont
  {de~Vasconcellos}}, \bibinfo {author} {\bibfnamefont {D.~P.}\ \bibnamefont
  {de~Lara}}, \bibinfo {author} {\bibfnamefont {R.}~\bibnamefont
  {Bratschitsch}}, \bibinfo {author} {\bibfnamefont {M.}~\bibnamefont
  {Rohlfing}},\ and\ \bibinfo {author} {\bibfnamefont {A.}~\bibnamefont
  {Castellanos-Gomez}},\ }\bibfield  {title} {\bibinfo {title} {Biaxial strain
  tuning of the optical properties of single-layer transition metal
  dichalcogenides},\ }\href@noop {} {\bibfield  {journal} {\bibinfo  {journal}
  {npj 2D Materials and Applications}\ }\textbf {\bibinfo {volume} {1}},\
  \bibinfo {pages} {1} (\bibinfo {year} {2017})}\BibitemShut {NoStop}%
\bibitem [{\citenamefont {Jadczak}\ \emph {et~al.}(2017)\citenamefont
  {Jadczak}, \citenamefont {Kutrowska-Girzycka}, \citenamefont
  {Kapu{\'s}ci{\'n}ski}, \citenamefont {Huang}, \citenamefont {W{\'o}js},\ and\
  \citenamefont {Bryja}}]{Jadczak2017}%
  \BibitemOpen
  \bibfield  {author} {\bibinfo {author} {\bibfnamefont {J.}~\bibnamefont
  {Jadczak}}, \bibinfo {author} {\bibfnamefont {J.}~\bibnamefont
  {Kutrowska-Girzycka}}, \bibinfo {author} {\bibfnamefont {P.}~\bibnamefont
  {Kapu{\'s}ci{\'n}ski}}, \bibinfo {author} {\bibfnamefont {Y.}~\bibnamefont
  {Huang}}, \bibinfo {author} {\bibfnamefont {A.}~\bibnamefont {W{\'o}js}},\
  and\ \bibinfo {author} {\bibfnamefont {z.}~\bibnamefont {Bryja}},\ }\bibfield
   {title} {\bibinfo {title} {Probing of free and localized excitons and trions
  in atomically thin {WSe}$_2$, {WS}$_2$, {MoSe}$_2$ and {MoS}$_2$ in
  photoluminescence and reflectivity experiments},\ }\href@noop {} {\bibfield
  {journal} {\bibinfo  {journal} {Nanotechnology}\ }\textbf {\bibinfo {volume}
  {28}},\ \bibinfo {pages} {395702} (\bibinfo {year} {2017})}\BibitemShut
  {NoStop}%
\bibitem [{\citenamefont {Koperski}\ \emph {et~al.}(2017)\citenamefont
  {Koperski}, \citenamefont {Molas}, \citenamefont {Arora}, \citenamefont
  {Nogajewski}, \citenamefont {Slobodeniuk}, \citenamefont {Faugeras},\ and\
  \citenamefont {Potemski}}]{Koperski2017}%
  \BibitemOpen
  \bibfield  {author} {\bibinfo {author} {\bibfnamefont {M.}~\bibnamefont
  {Koperski}}, \bibinfo {author} {\bibfnamefont {M.~R.}\ \bibnamefont {Molas}},
  \bibinfo {author} {\bibfnamefont {A.}~\bibnamefont {Arora}}, \bibinfo
  {author} {\bibfnamefont {K.}~\bibnamefont {Nogajewski}}, \bibinfo {author}
  {\bibfnamefont {A.~O.}\ \bibnamefont {Slobodeniuk}}, \bibinfo {author}
  {\bibfnamefont {C.}~\bibnamefont {Faugeras}},\ and\ \bibinfo {author}
  {\bibfnamefont {M.}~\bibnamefont {Potemski}},\ }\bibfield  {title} {\bibinfo
  {title} {Optical properties of atomically thin transition metal
  dichalcogenides: observations and puzzles},\ }\href@noop {} {\bibfield
  {journal} {\bibinfo  {journal} {Nanophotonics}\ }\textbf {\bibinfo {volume}
  {6}},\ \bibinfo {pages} {1289} (\bibinfo {year} {2017})}\BibitemShut
  {NoStop}%
\bibitem [{\citenamefont {Xu}\ \emph {et~al.}(2021)\citenamefont {Xu},
  \citenamefont {Horn}, \citenamefont {Zhu}, \citenamefont {Tang},
  \citenamefont {Ma}, \citenamefont {Li}, \citenamefont {Liu}, \citenamefont
  {Watanabe}, \citenamefont {Taniguchi}, \citenamefont {Hone} \emph
  {et~al.}}]{Xu2021}%
  \BibitemOpen
  \bibfield  {author} {\bibinfo {author} {\bibfnamefont {Y.}~\bibnamefont
  {Xu}}, \bibinfo {author} {\bibfnamefont {C.}~\bibnamefont {Horn}}, \bibinfo
  {author} {\bibfnamefont {J.}~\bibnamefont {Zhu}}, \bibinfo {author}
  {\bibfnamefont {Y.}~\bibnamefont {Tang}}, \bibinfo {author} {\bibfnamefont
  {L.}~\bibnamefont {Ma}}, \bibinfo {author} {\bibfnamefont {L.}~\bibnamefont
  {Li}}, \bibinfo {author} {\bibfnamefont {S.}~\bibnamefont {Liu}}, \bibinfo
  {author} {\bibfnamefont {K.}~\bibnamefont {Watanabe}}, \bibinfo {author}
  {\bibfnamefont {T.}~\bibnamefont {Taniguchi}}, \bibinfo {author}
  {\bibfnamefont {J.~C.}\ \bibnamefont {Hone}}, \emph {et~al.},\ }\bibfield
  {title} {\bibinfo {title} {Creation of moir{\'e} bands in a monolayer
  semiconductor by spatially periodic dielectric screening},\ }\href@noop {}
  {\bibfield  {journal} {\bibinfo  {journal} {Nat. Mat.}\ ,\ \bibinfo {pages}
  {1}} (\bibinfo {year} {2021})}\BibitemShut {NoStop}%
\bibitem [{\citenamefont {Polman}\ \emph {et~al.}(2019)\citenamefont {Polman},
  \citenamefont {Kociak},\ and\ \citenamefont {de~Abajo}}]{Polman2019}%
  \BibitemOpen
  \bibfield  {author} {\bibinfo {author} {\bibfnamefont {A.}~\bibnamefont
  {Polman}}, \bibinfo {author} {\bibfnamefont {M.}~\bibnamefont {Kociak}},\
  and\ \bibinfo {author} {\bibfnamefont {F.~J.~G.}\ \bibnamefont {de~Abajo}},\
  }\bibfield  {title} {\bibinfo {title} {Electron-beam spectroscopy for
  nanophotonics},\ }\href@noop {} {\bibfield  {journal} {\bibinfo  {journal}
  {Nat. Mat.}\ }\textbf {\bibinfo {volume} {18}},\ \bibinfo {pages} {1158}
  (\bibinfo {year} {2019})}\BibitemShut {NoStop}%
\bibitem [{\citenamefont {Zheng}\ \emph {et~al.}(2017)\citenamefont {Zheng},
  \citenamefont {So}, \citenamefont {Liu}, \citenamefont {Liu}, \citenamefont
  {Zheludev},\ and\ \citenamefont {Fan}}]{Zheng2017}%
  \BibitemOpen
  \bibfield  {author} {\bibinfo {author} {\bibfnamefont {S.}~\bibnamefont
  {Zheng}}, \bibinfo {author} {\bibfnamefont {J.-K.}\ \bibnamefont {So}},
  \bibinfo {author} {\bibfnamefont {F.}~\bibnamefont {Liu}}, \bibinfo {author}
  {\bibfnamefont {Z.}~\bibnamefont {Liu}}, \bibinfo {author} {\bibfnamefont
  {N.}~\bibnamefont {Zheludev}},\ and\ \bibinfo {author} {\bibfnamefont
  {H.~J.}\ \bibnamefont {Fan}},\ }\bibfield  {title} {\bibinfo {title} {Giant
  enhancement of cathodoluminescence of monolayer transitional metal
  dichalcogenides semiconductors},\ }\href@noop {} {\bibfield  {journal}
  {\bibinfo  {journal} {Nano Lett.}\ }\textbf {\bibinfo {volume} {17}},\
  \bibinfo {pages} {6475} (\bibinfo {year} {2017})}\BibitemShut {NoStop}%
\bibitem [{\citenamefont {Nayak}\ \emph {et~al.}(2019)\citenamefont {Nayak},
  \citenamefont {Lisi}, \citenamefont {Liu}, \citenamefont {Jakubczyk},
  \citenamefont {Stepanov}, \citenamefont {Donatini}, \citenamefont {Watanabe},
  \citenamefont {Taniguchi}, \citenamefont {Bid}, \citenamefont {Kasprzak}
  \emph {et~al.}}]{Nayak2019}%
  \BibitemOpen
  \bibfield  {author} {\bibinfo {author} {\bibfnamefont {G.}~\bibnamefont
  {Nayak}}, \bibinfo {author} {\bibfnamefont {S.}~\bibnamefont {Lisi}},
  \bibinfo {author} {\bibfnamefont {W.}~\bibnamefont {Liu}}, \bibinfo {author}
  {\bibfnamefont {T.}~\bibnamefont {Jakubczyk}}, \bibinfo {author}
  {\bibfnamefont {P.}~\bibnamefont {Stepanov}}, \bibinfo {author}
  {\bibfnamefont {F.}~\bibnamefont {Donatini}}, \bibinfo {author}
  {\bibfnamefont {K.}~\bibnamefont {Watanabe}}, \bibinfo {author}
  {\bibfnamefont {T.}~\bibnamefont {Taniguchi}}, \bibinfo {author}
  {\bibfnamefont {A.}~\bibnamefont {Bid}}, \bibinfo {author} {\bibfnamefont
  {J.}~\bibnamefont {Kasprzak}}, \emph {et~al.},\ }\bibfield  {title} {\bibinfo
  {title} {Cathodoluminescence enhancement and quenching in type-i van der
  waals heterostructures: Cleanliness of the interfaces and defect creation},\
  }\href@noop {} {\bibfield  {journal} {\bibinfo  {journal} {Phys. Rev.
  Materials}\ }\textbf {\bibinfo {volume} {3}},\ \bibinfo {pages} {114001}
  (\bibinfo {year} {2019})}\BibitemShut {NoStop}%
\bibitem [{\citenamefont {Singh}\ \emph {et~al.}(2020)\citenamefont {Singh},
  \citenamefont {Lee},\ and\ \citenamefont {Grade{\v{c}}ak}}]{Singh2020}%
  \BibitemOpen
  \bibfield  {author} {\bibinfo {author} {\bibfnamefont {A.}~\bibnamefont
  {Singh}}, \bibinfo {author} {\bibfnamefont {H.~Y.}\ \bibnamefont {Lee}},\
  and\ \bibinfo {author} {\bibfnamefont {S.}~\bibnamefont {Grade{\v{c}}ak}},\
  }\bibfield  {title} {\bibinfo {title} {Direct optical-structure correlation
  in atomically thin dichalcogenides and heterostructures},\ }\href@noop {}
  {\bibfield  {journal} {\bibinfo  {journal} {Nano Res.}\ }\textbf {\bibinfo
  {volume} {13}},\ \bibinfo {pages} {1} (\bibinfo {year} {2020})}\BibitemShut
  {NoStop}%
\bibitem [{\citenamefont {Tizei}\ \emph {et~al.}(2015)\citenamefont {Tizei},
  \citenamefont {Lin}, \citenamefont {Mukai}, \citenamefont {Sawada},
  \citenamefont {Lu}, \citenamefont {Li}, \citenamefont {Kimoto},\ and\
  \citenamefont {Suenaga}}]{Tizei2015}%
  \BibitemOpen
  \bibfield  {author} {\bibinfo {author} {\bibfnamefont {L.~H.}\ \bibnamefont
  {Tizei}}, \bibinfo {author} {\bibfnamefont {Y.-C.}\ \bibnamefont {Lin}},
  \bibinfo {author} {\bibfnamefont {M.}~\bibnamefont {Mukai}}, \bibinfo
  {author} {\bibfnamefont {H.}~\bibnamefont {Sawada}}, \bibinfo {author}
  {\bibfnamefont {A.-Y.}\ \bibnamefont {Lu}}, \bibinfo {author} {\bibfnamefont
  {L.-J.}\ \bibnamefont {Li}}, \bibinfo {author} {\bibfnamefont
  {K.}~\bibnamefont {Kimoto}},\ and\ \bibinfo {author} {\bibfnamefont
  {K.}~\bibnamefont {Suenaga}},\ }\bibfield  {title} {\bibinfo {title} {Exciton
  mapping at subwavelength scales in two-dimensional materials},\ }\href@noop
  {} {\bibfield  {journal} {\bibinfo  {journal} {Phys. Rev. Lett.}\ }\textbf
  {\bibinfo {volume} {114}},\ \bibinfo {pages} {107601} (\bibinfo {year}
  {2015})}\BibitemShut {NoStop}%
\bibitem [{\citenamefont {Habenicht}\ \emph {et~al.}(2015)\citenamefont
  {Habenicht}, \citenamefont {Knupfer},\ and\ \citenamefont
  {B{\"u}chner}}]{Habenicht2015}%
  \BibitemOpen
  \bibfield  {author} {\bibinfo {author} {\bibfnamefont {C.}~\bibnamefont
  {Habenicht}}, \bibinfo {author} {\bibfnamefont {M.}~\bibnamefont {Knupfer}},\
  and\ \bibinfo {author} {\bibfnamefont {B.}~\bibnamefont {B{\"u}chner}},\
  }\bibfield  {title} {\bibinfo {title} {Investigation of the dispersion and
  the effective masses of excitons in bulk {2H-MoS}$_2$ using transition
  electron energy-loss spectroscopy},\ }\href@noop {} {\bibfield  {journal}
  {\bibinfo  {journal} {Phys. Rev. B}\ }\textbf {\bibinfo {volume} {91}},\
  \bibinfo {pages} {245203} (\bibinfo {year} {2015})}\BibitemShut {NoStop}%
\bibitem [{\citenamefont {Nerl}\ \emph {et~al.}(2017)\citenamefont {Nerl},
  \citenamefont {Winther}, \citenamefont {Hage}, \citenamefont {Thygesen},
  \citenamefont {Houben}, \citenamefont {Backes}, \citenamefont {Coleman},
  \citenamefont {Ramasse},\ and\ \citenamefont {Nicolosi}}]{Nerl2017}%
  \BibitemOpen
  \bibfield  {author} {\bibinfo {author} {\bibfnamefont {H.~C.}\ \bibnamefont
  {Nerl}}, \bibinfo {author} {\bibfnamefont {K.~T.}\ \bibnamefont {Winther}},
  \bibinfo {author} {\bibfnamefont {F.~S.}\ \bibnamefont {Hage}}, \bibinfo
  {author} {\bibfnamefont {K.~S.}\ \bibnamefont {Thygesen}}, \bibinfo {author}
  {\bibfnamefont {L.}~\bibnamefont {Houben}}, \bibinfo {author} {\bibfnamefont
  {C.}~\bibnamefont {Backes}}, \bibinfo {author} {\bibfnamefont {J.~N.}\
  \bibnamefont {Coleman}}, \bibinfo {author} {\bibfnamefont {Q.~M.}\
  \bibnamefont {Ramasse}},\ and\ \bibinfo {author} {\bibfnamefont
  {V.}~\bibnamefont {Nicolosi}},\ }\bibfield  {title} {\bibinfo {title}
  {Probing the local nature of excitons and plasmons in few-layer {MoS}$_2$},\
  }\href@noop {} {\bibfield  {journal} {\bibinfo  {journal} {npj 2D Materials
  and Applications}\ }\textbf {\bibinfo {volume} {1}},\ \bibinfo {pages} {1}
  (\bibinfo {year} {2017})}\BibitemShut {NoStop}%
\bibitem [{\citenamefont {Hong}\ \emph {et~al.}(2020)\citenamefont {Hong},
  \citenamefont {Senga}, \citenamefont {Pichler},\ and\ \citenamefont
  {Suenaga}}]{Hong2020}%
  \BibitemOpen
  \bibfield  {author} {\bibinfo {author} {\bibfnamefont {J.}~\bibnamefont
  {Hong}}, \bibinfo {author} {\bibfnamefont {R.}~\bibnamefont {Senga}},
  \bibinfo {author} {\bibfnamefont {T.}~\bibnamefont {Pichler}},\ and\ \bibinfo
  {author} {\bibfnamefont {K.}~\bibnamefont {Suenaga}},\ }\bibfield  {title}
  {\bibinfo {title} {Probing exciton dispersions of freestanding monolayer
  {WSe}$_{2}$ by momentum-resolved electron energy-loss spectroscopy},\ }\href
  {https://doi.org/10.1103/PhysRevLett.124.087401} {\bibfield  {journal}
  {\bibinfo  {journal} {Phys. Rev. Lett.}\ }\textbf {\bibinfo {volume} {124}},\
  \bibinfo {pages} {087401} (\bibinfo {year} {2020})}\BibitemShut {NoStop}%
\bibitem [{\citenamefont {Peña~Román}\ \emph {et~al.}(2020)\citenamefont
  {Peña~Román}, \citenamefont {Auad}, \citenamefont {Grasso}, \citenamefont
  {Alvarez}, \citenamefont {Barcelos},\ and\ \citenamefont
  {Zagonel}}]{PenaRoman2020}%
  \BibitemOpen
  \bibfield  {author} {\bibinfo {author} {\bibfnamefont {R.~J.}\ \bibnamefont
  {Peña~Román}}, \bibinfo {author} {\bibfnamefont {Y.}~\bibnamefont {Auad}},
  \bibinfo {author} {\bibfnamefont {L.}~\bibnamefont {Grasso}}, \bibinfo
  {author} {\bibfnamefont {F.}~\bibnamefont {Alvarez}}, \bibinfo {author}
  {\bibfnamefont {I.~D.}\ \bibnamefont {Barcelos}},\ and\ \bibinfo {author}
  {\bibfnamefont {L.~F.}\ \bibnamefont {Zagonel}},\ }\bibfield  {title}
  {\bibinfo {title} {Tunneling-current-induced local excitonic luminescence in
  p-doped {WSe}$_2$ monolayers},\ }\href {https://doi.org/10.1039/D0NR03400B}
  {\bibfield  {journal} {\bibinfo  {journal} {Nanoscale}\ }\textbf {\bibinfo
  {volume} {12}},\ \bibinfo {pages} {13460} (\bibinfo {year}
  {2020})}\BibitemShut {NoStop}%
\bibitem [{\citenamefont {Schuler}\ \emph {et~al.}(2020)\citenamefont
  {Schuler}, \citenamefont {Cochrane}, \citenamefont {Kastl}, \citenamefont
  {Barnard}, \citenamefont {Wong}, \citenamefont {Borys}, \citenamefont
  {Schwartzberg}, \citenamefont {Ogletree}, \citenamefont {de~Abajo},\ and\
  \citenamefont {Weber-Bargioni}}]{Schuler2020}%
  \BibitemOpen
  \bibfield  {author} {\bibinfo {author} {\bibfnamefont {B.}~\bibnamefont
  {Schuler}}, \bibinfo {author} {\bibfnamefont {K.~A.}\ \bibnamefont
  {Cochrane}}, \bibinfo {author} {\bibfnamefont {C.}~\bibnamefont {Kastl}},
  \bibinfo {author} {\bibfnamefont {E.~S.}\ \bibnamefont {Barnard}}, \bibinfo
  {author} {\bibfnamefont {E.}~\bibnamefont {Wong}}, \bibinfo {author}
  {\bibfnamefont {N.~J.}\ \bibnamefont {Borys}}, \bibinfo {author}
  {\bibfnamefont {A.~M.}\ \bibnamefont {Schwartzberg}}, \bibinfo {author}
  {\bibfnamefont {D.~F.}\ \bibnamefont {Ogletree}}, \bibinfo {author}
  {\bibfnamefont {F.~J.~G.}\ \bibnamefont {de~Abajo}},\ and\ \bibinfo {author}
  {\bibfnamefont {A.}~\bibnamefont {Weber-Bargioni}},\ }\bibfield  {title}
  {\bibinfo {title} {Electrically driven photon emission from individual atomic
  defects in monolayer {WS}$_2$},\ }\bibfield  {journal} {\bibinfo  {journal}
  {Science Advances}\ }\textbf {\bibinfo {volume} {6}},\ \href
  {https://doi.org/10.1126/sciadv.abb5988} {10.1126/sciadv.abb5988} (\bibinfo
  {year} {2020})\BibitemShut {NoStop}%
\bibitem [{\citenamefont {Mahfoud}\ \emph {et~al.}(2013)\citenamefont
  {Mahfoud}, \citenamefont {Dijksman}, \citenamefont {Javaux}, \citenamefont
  {Bassoul}, \citenamefont {Baudrion}, \citenamefont {Plain}, \citenamefont
  {Dubertret},\ and\ \citenamefont {Kociak}}]{Mahfoud2013}%
  \BibitemOpen
  \bibfield  {author} {\bibinfo {author} {\bibfnamefont {Z.}~\bibnamefont
  {Mahfoud}}, \bibinfo {author} {\bibfnamefont {A.~T.}\ \bibnamefont
  {Dijksman}}, \bibinfo {author} {\bibfnamefont {C.}~\bibnamefont {Javaux}},
  \bibinfo {author} {\bibfnamefont {P.}~\bibnamefont {Bassoul}}, \bibinfo
  {author} {\bibfnamefont {A.-L.}\ \bibnamefont {Baudrion}}, \bibinfo {author}
  {\bibfnamefont {J.}~\bibnamefont {Plain}}, \bibinfo {author} {\bibfnamefont
  {B.}~\bibnamefont {Dubertret}},\ and\ \bibinfo {author} {\bibfnamefont
  {M.}~\bibnamefont {Kociak}},\ }\bibfield  {title} {\bibinfo {title}
  {Cathodoluminescence in a scanning transmission electron microscope: A
  nanometer-scale counterpart of photoluminescence for the study of ii--vi
  quantum dots},\ }\href@noop {} {\bibfield  {journal} {\bibinfo  {journal} {J.
  of Phys. Chem. Lett.}\ }\textbf {\bibinfo {volume} {4}},\ \bibinfo {pages}
  {4090} (\bibinfo {year} {2013})}\BibitemShut {NoStop}%
\bibitem [{\citenamefont {Kolesnichenko}\ \emph {et~al.}(2020)\citenamefont
  {Kolesnichenko}, \citenamefont {Zhang}, \citenamefont {Yun}, \citenamefont
  {Zheng}, \citenamefont {Fuhrer},\ and\ \citenamefont
  {Davis}}]{Kolesnichenko2020}%
  \BibitemOpen
  \bibfield  {author} {\bibinfo {author} {\bibfnamefont {P.~V.}\ \bibnamefont
  {Kolesnichenko}}, \bibinfo {author} {\bibfnamefont {Q.}~\bibnamefont
  {Zhang}}, \bibinfo {author} {\bibfnamefont {T.}~\bibnamefont {Yun}}, \bibinfo
  {author} {\bibfnamefont {C.}~\bibnamefont {Zheng}}, \bibinfo {author}
  {\bibfnamefont {M.~S.}\ \bibnamefont {Fuhrer}},\ and\ \bibinfo {author}
  {\bibfnamefont {J.~A.}\ \bibnamefont {Davis}},\ }\bibfield  {title} {\bibinfo
  {title} {Disentangling the effects of doping, strain and disorder in
  monolayer {WS}$_2$ by optical spectroscopy},\ }\href@noop {} {\bibfield
  {journal} {\bibinfo  {journal} {2D Materials}\ }\textbf {\bibinfo {volume}
  {7}},\ \bibinfo {pages} {025008} (\bibinfo {year} {2020})}\BibitemShut
  {NoStop}%
\bibitem [{\citenamefont {Niehues}\ \emph {et~al.}(2020)\citenamefont
  {Niehues}, \citenamefont {Marauhn}, \citenamefont {Deilmann}, \citenamefont
  {Wigger}, \citenamefont {Schmidt}, \citenamefont {Arora}, \citenamefont
  {de~Vasconcellos}, \citenamefont {Rohlfing},\ and\ \citenamefont
  {Bratschitsch}}]{Niehues2020}%
  \BibitemOpen
  \bibfield  {author} {\bibinfo {author} {\bibfnamefont {I.}~\bibnamefont
  {Niehues}}, \bibinfo {author} {\bibfnamefont {P.}~\bibnamefont {Marauhn}},
  \bibinfo {author} {\bibfnamefont {T.}~\bibnamefont {Deilmann}}, \bibinfo
  {author} {\bibfnamefont {D.}~\bibnamefont {Wigger}}, \bibinfo {author}
  {\bibfnamefont {R.}~\bibnamefont {Schmidt}}, \bibinfo {author} {\bibfnamefont
  {A.}~\bibnamefont {Arora}}, \bibinfo {author} {\bibfnamefont {S.~M.}\
  \bibnamefont {de~Vasconcellos}}, \bibinfo {author} {\bibfnamefont
  {M.}~\bibnamefont {Rohlfing}},\ and\ \bibinfo {author} {\bibfnamefont
  {R.}~\bibnamefont {Bratschitsch}},\ }\bibfield  {title} {\bibinfo {title}
  {Strain tuning of the stokes shift in atomically thin semiconductors},\
  }\href@noop {} {\bibfield  {journal} {\bibinfo  {journal} {Nanoscale}\
  }\textbf {\bibinfo {volume} {12}},\ \bibinfo {pages} {20786} (\bibinfo {year}
  {2020})}\BibitemShut {NoStop}%
\bibitem [{\citenamefont {Hambach}(2010)}]{Hambach2010}%
  \BibitemOpen
  \bibfield  {author} {\bibinfo {author} {\bibfnamefont {R.}~\bibnamefont
  {Hambach}},\ }\emph {\bibinfo {title} {Theory and ab-initio calculations of
  collective excitations in nanostructures: towards spatially-resolved EELS}},\
  \href@noop {} {Ph.D. thesis} (\bibinfo {year} {2010})\BibitemShut {NoStop}%
\bibitem [{\citenamefont {Kociak}\ and\ \citenamefont
  {Zagonel}(2017)}]{Kociak2017}%
  \BibitemOpen
  \bibfield  {author} {\bibinfo {author} {\bibfnamefont {M.}~\bibnamefont
  {Kociak}}\ and\ \bibinfo {author} {\bibfnamefont {L.}~\bibnamefont
  {Zagonel}},\ }\bibfield  {title} {\bibinfo {title} {Cathodoluminescence in
  the scanning transmission electron microscope},\ }\href@noop {} {\bibfield
  {journal} {\bibinfo  {journal} {Ultramicroscopy}\ }\textbf {\bibinfo {volume}
  {176}},\ \bibinfo {pages} {112} (\bibinfo {year} {2017})}\BibitemShut
  {NoStop}%
\bibitem [{\citenamefont {Carvalho}\ \emph {et~al.}(2013)\citenamefont
  {Carvalho}, \citenamefont {Ribeiro},\ and\ \citenamefont
  {Neto}}]{Carvalho2013}%
  \BibitemOpen
  \bibfield  {author} {\bibinfo {author} {\bibfnamefont {A.}~\bibnamefont
  {Carvalho}}, \bibinfo {author} {\bibfnamefont {R.}~\bibnamefont {Ribeiro}},\
  and\ \bibinfo {author} {\bibfnamefont {A.~C.}\ \bibnamefont {Neto}},\
  }\bibfield  {title} {\bibinfo {title} {Band nesting and the optical response
  of two-dimensional semiconducting transition metal dichalcogenides},\
  }\href@noop {} {\bibfield  {journal} {\bibinfo  {journal} {Phys. Rev. B}\
  }\textbf {\bibinfo {volume} {88}},\ \bibinfo {pages} {115205} (\bibinfo
  {year} {2013})}\BibitemShut {NoStop}%
\bibitem [{\citenamefont {Bourrellier}\ \emph {et~al.}(2016)\citenamefont
  {Bourrellier}, \citenamefont {Meuret}, \citenamefont {Tararan}, \citenamefont
  {St{\'e}phan}, \citenamefont {Kociak}, \citenamefont {Tizei},\ and\
  \citenamefont {Zobelli}}]{Bourrellier2016}%
  \BibitemOpen
  \bibfield  {author} {\bibinfo {author} {\bibfnamefont {R.}~\bibnamefont
  {Bourrellier}}, \bibinfo {author} {\bibfnamefont {S.}~\bibnamefont {Meuret}},
  \bibinfo {author} {\bibfnamefont {A.}~\bibnamefont {Tararan}}, \bibinfo
  {author} {\bibfnamefont {O.}~\bibnamefont {St{\'e}phan}}, \bibinfo {author}
  {\bibfnamefont {M.}~\bibnamefont {Kociak}}, \bibinfo {author} {\bibfnamefont
  {L.~H.~G.}\ \bibnamefont {Tizei}},\ and\ \bibinfo {author} {\bibfnamefont
  {A.}~\bibnamefont {Zobelli}},\ }\bibfield  {title} {\bibinfo {title} {Bright
  {UV} single photon emission at point defects in h-{BN}},\ }\href@noop {}
  {\bibfield  {journal} {\bibinfo  {journal} {Nano Lett.}\ }\textbf {\bibinfo
  {volume} {16}},\ \bibinfo {pages} {4317} (\bibinfo {year}
  {2016})}\BibitemShut {NoStop}%
\bibitem [{\citenamefont {Haigh}\ \emph {et~al.}(2012)\citenamefont {Haigh},
  \citenamefont {Gholinia}, \citenamefont {Jalil}, \citenamefont {Romani},
  \citenamefont {Britnell}, \citenamefont {Elias}, \citenamefont {Novoselov},
  \citenamefont {Ponomarenko}, \citenamefont {Geim},\ and\ \citenamefont
  {Gorbachev}}]{Haigh2012}%
  \BibitemOpen
  \bibfield  {author} {\bibinfo {author} {\bibfnamefont {S.}~\bibnamefont
  {Haigh}}, \bibinfo {author} {\bibfnamefont {A.}~\bibnamefont {Gholinia}},
  \bibinfo {author} {\bibfnamefont {R.}~\bibnamefont {Jalil}}, \bibinfo
  {author} {\bibfnamefont {S.}~\bibnamefont {Romani}}, \bibinfo {author}
  {\bibfnamefont {L.}~\bibnamefont {Britnell}}, \bibinfo {author}
  {\bibfnamefont {D.}~\bibnamefont {Elias}}, \bibinfo {author} {\bibfnamefont
  {K.}~\bibnamefont {Novoselov}}, \bibinfo {author} {\bibfnamefont
  {L.}~\bibnamefont {Ponomarenko}}, \bibinfo {author} {\bibfnamefont
  {A.}~\bibnamefont {Geim}},\ and\ \bibinfo {author} {\bibfnamefont
  {R.}~\bibnamefont {Gorbachev}},\ }\bibfield  {title} {\bibinfo {title}
  {Cross-sectional imaging of individual layers and buried interfaces of
  graphene-based heterostructures and superlattices},\ }\href@noop {}
  {\bibfield  {journal} {\bibinfo  {journal} {Nature materials}\ }\textbf
  {\bibinfo {volume} {11}},\ \bibinfo {pages} {764} (\bibinfo {year}
  {2012})}\BibitemShut {NoStop}%
\bibitem [{\citenamefont {Teran}\ \emph {et~al.}(2005)\citenamefont {Teran},
  \citenamefont {Eaves}, \citenamefont {Mansouri}, \citenamefont {Buhmann},
  \citenamefont {Maude}, \citenamefont {Potemski}, \citenamefont {Henini},\
  and\ \citenamefont {Hill}}]{Teran2005}%
  \BibitemOpen
  \bibfield  {author} {\bibinfo {author} {\bibfnamefont {F.~J.}\ \bibnamefont
  {Teran}}, \bibinfo {author} {\bibfnamefont {L.}~\bibnamefont {Eaves}},
  \bibinfo {author} {\bibfnamefont {L.}~\bibnamefont {Mansouri}}, \bibinfo
  {author} {\bibfnamefont {H.}~\bibnamefont {Buhmann}}, \bibinfo {author}
  {\bibfnamefont {D.~K.}\ \bibnamefont {Maude}}, \bibinfo {author}
  {\bibfnamefont {M.}~\bibnamefont {Potemski}}, \bibinfo {author}
  {\bibfnamefont {M.}~\bibnamefont {Henini}},\ and\ \bibinfo {author}
  {\bibfnamefont {G.}~\bibnamefont {Hill}},\ }\bibfield  {title} {\bibinfo
  {title} {Trion formation in narrow {GaAs} quantum well structures},\ }\href
  {https://doi.org/10.1103/PhysRevB.71.161309} {\bibfield  {journal} {\bibinfo
  {journal} {Phys. Rev. B}\ }\textbf {\bibinfo {volume} {71}},\ \bibinfo
  {pages} {161309} (\bibinfo {year} {2005})}\BibitemShut {NoStop}%
\bibitem [{\citenamefont {Peimyoo}\ \emph {et~al.}(2014)\citenamefont
  {Peimyoo}, \citenamefont {Yang}, \citenamefont {Shang}, \citenamefont {Shen},
  \citenamefont {Wang},\ and\ \citenamefont {Yu}}]{Peimyoo2014}%
  \BibitemOpen
  \bibfield  {author} {\bibinfo {author} {\bibfnamefont {N.}~\bibnamefont
  {Peimyoo}}, \bibinfo {author} {\bibfnamefont {W.}~\bibnamefont {Yang}},
  \bibinfo {author} {\bibfnamefont {J.}~\bibnamefont {Shang}}, \bibinfo
  {author} {\bibfnamefont {X.}~\bibnamefont {Shen}}, \bibinfo {author}
  {\bibfnamefont {Y.}~\bibnamefont {Wang}},\ and\ \bibinfo {author}
  {\bibfnamefont {T.}~\bibnamefont {Yu}},\ }\bibfield  {title} {\bibinfo
  {title} {Chemically driven tunable light emission of charged and neutral
  excitons in monolayer {WS}$_2$},\ }\href@noop {} {\bibfield  {journal}
  {\bibinfo  {journal} {ACS Nano}\ }\textbf {\bibinfo {volume} {8}},\ \bibinfo
  {pages} {11320} (\bibinfo {year} {2014})}\BibitemShut {NoStop}%
\bibitem [{\citenamefont {Neumann}\ \emph {et~al.}(2018)\citenamefont
  {Neumann}, \citenamefont {Lindlau}, \citenamefont {Nutz}, \citenamefont
  {Mohite}, \citenamefont {Yamaguchi},\ and\ \citenamefont
  {H\"ogele}}]{Neumann2018}%
  \BibitemOpen
  \bibfield  {author} {\bibinfo {author} {\bibfnamefont {A.}~\bibnamefont
  {Neumann}}, \bibinfo {author} {\bibfnamefont {J.}~\bibnamefont {Lindlau}},
  \bibinfo {author} {\bibfnamefont {M.}~\bibnamefont {Nutz}}, \bibinfo {author}
  {\bibfnamefont {A.~D.}\ \bibnamefont {Mohite}}, \bibinfo {author}
  {\bibfnamefont {H.}~\bibnamefont {Yamaguchi}},\ and\ \bibinfo {author}
  {\bibfnamefont {A.}~\bibnamefont {H\"ogele}},\ }\bibfield  {title} {\bibinfo
  {title} {Signatures of defect-localized charged excitons in the
  photoluminescence of monolayer molybdenum disulfide},\ }\href
  {https://doi.org/10.1103/PhysRevMaterials.2.124003} {\bibfield  {journal}
  {\bibinfo  {journal} {Phys. Rev. Materials}\ }\textbf {\bibinfo {volume}
  {2}},\ \bibinfo {pages} {124003} (\bibinfo {year} {2018})}\BibitemShut
  {NoStop}%
\bibitem [{\citenamefont {Lin}\ \emph {et~al.}(2018)\citenamefont {Lin},
  \citenamefont {Li}, \citenamefont {Komsa}, \citenamefont {Chang},
  \citenamefont {Krasheninnikov}, \citenamefont {Eda},\ and\ \citenamefont
  {Suenaga}}]{Lin2018}%
  \BibitemOpen
  \bibfield  {author} {\bibinfo {author} {\bibfnamefont {Y.-C.}\ \bibnamefont
  {Lin}}, \bibinfo {author} {\bibfnamefont {S.}~\bibnamefont {Li}}, \bibinfo
  {author} {\bibfnamefont {H.-P.}\ \bibnamefont {Komsa}}, \bibinfo {author}
  {\bibfnamefont {L.-J.}\ \bibnamefont {Chang}}, \bibinfo {author}
  {\bibfnamefont {A.~V.}\ \bibnamefont {Krasheninnikov}}, \bibinfo {author}
  {\bibfnamefont {G.}~\bibnamefont {Eda}},\ and\ \bibinfo {author}
  {\bibfnamefont {K.}~\bibnamefont {Suenaga}},\ }\bibfield  {title} {\bibinfo
  {title} {Revealing the atomic defects of {WS}$_2$ governing its distinct
  optical emissions},\ }\href@noop {} {\bibfield  {journal} {\bibinfo
  {journal} {Adv. Func. Mat.}\ }\textbf {\bibinfo {volume} {28}},\ \bibinfo
  {pages} {1704210} (\bibinfo {year} {2018})}\BibitemShut {NoStop}%
\bibitem [{\citenamefont {Kobayashi}\ \emph {et~al.}(2015)\citenamefont
  {Kobayashi}, \citenamefont {Sasaki}, \citenamefont {Mori}, \citenamefont
  {Hibino}, \citenamefont {Liu}, \citenamefont {Watanabe}, \citenamefont
  {Taniguchi}, \citenamefont {Suenaga}, \citenamefont {Maniwa},\ and\
  \citenamefont {Miyata}}]{Kobayashi2015}%
  \BibitemOpen
  \bibfield  {author} {\bibinfo {author} {\bibfnamefont {Y.}~\bibnamefont
  {Kobayashi}}, \bibinfo {author} {\bibfnamefont {S.}~\bibnamefont {Sasaki}},
  \bibinfo {author} {\bibfnamefont {S.}~\bibnamefont {Mori}}, \bibinfo {author}
  {\bibfnamefont {H.}~\bibnamefont {Hibino}}, \bibinfo {author} {\bibfnamefont
  {Z.}~\bibnamefont {Liu}}, \bibinfo {author} {\bibfnamefont {K.}~\bibnamefont
  {Watanabe}}, \bibinfo {author} {\bibfnamefont {T.}~\bibnamefont {Taniguchi}},
  \bibinfo {author} {\bibfnamefont {K.}~\bibnamefont {Suenaga}}, \bibinfo
  {author} {\bibfnamefont {Y.}~\bibnamefont {Maniwa}},\ and\ \bibinfo {author}
  {\bibfnamefont {Y.}~\bibnamefont {Miyata}},\ }\bibfield  {title} {\bibinfo
  {title} {Growth and optical properties of high-quality monolayer {WS}$_2$ on
  graphite},\ }\href@noop {} {\bibfield  {journal} {\bibinfo  {journal} {ACS
  Nano}\ }\textbf {\bibinfo {volume} {9}},\ \bibinfo {pages} {4056} (\bibinfo
  {year} {2015})}\BibitemShut {NoStop}%
\bibitem [{\citenamefont {Harats}\ \emph {et~al.}(2020)\citenamefont {Harats},
  \citenamefont {Kirchhof}, \citenamefont {Qiao}, \citenamefont {Greben},\ and\
  \citenamefont {Bolotin}}]{Harats2020}%
  \BibitemOpen
  \bibfield  {author} {\bibinfo {author} {\bibfnamefont {M.~G.}\ \bibnamefont
  {Harats}}, \bibinfo {author} {\bibfnamefont {J.~N.}\ \bibnamefont
  {Kirchhof}}, \bibinfo {author} {\bibfnamefont {M.}~\bibnamefont {Qiao}},
  \bibinfo {author} {\bibfnamefont {K.}~\bibnamefont {Greben}},\ and\ \bibinfo
  {author} {\bibfnamefont {K.~I.}\ \bibnamefont {Bolotin}},\ }\bibfield
  {title} {\bibinfo {title} {Dynamics and efficient conversion of excitons to
  trions in non-uniformly strained monolayer {WS}$_ 2$},\ }\href@noop {}
  {\bibfield  {journal} {\bibinfo  {journal} {Nat. Photon.}\ ,\ \bibinfo
  {pages} {1}} (\bibinfo {year} {2020})}\BibitemShut {NoStop}%
\bibitem [{\citenamefont {Anger}\ \emph {et~al.}(2006)\citenamefont {Anger},
  \citenamefont {Bharadwaj},\ and\ \citenamefont {Novotny}}]{Anger2006}%
  \BibitemOpen
  \bibfield  {author} {\bibinfo {author} {\bibfnamefont {P.}~\bibnamefont
  {Anger}}, \bibinfo {author} {\bibfnamefont {P.}~\bibnamefont {Bharadwaj}},\
  and\ \bibinfo {author} {\bibfnamefont {L.}~\bibnamefont {Novotny}},\
  }\bibfield  {title} {\bibinfo {title} {Enhancement and quenching of
  single-molecule fluorescence},\ }\href@noop {} {\bibfield  {journal}
  {\bibinfo  {journal} {Phys. Rev. Lett.}\ }\textbf {\bibinfo {volume} {96}},\
  \bibinfo {pages} {113002} (\bibinfo {year} {2006})}\BibitemShut {NoStop}%
\bibitem [{\citenamefont {Tizei}\ \emph {et~al.}(2020)\citenamefont {Tizei},
  \citenamefont {Mkhitaryan}, \citenamefont {Louren{\c{c}}o-Martins},
  \citenamefont {Scarabelli}, \citenamefont {Watanabe}, \citenamefont
  {Taniguchi}, \citenamefont {Tenc{\'e}}, \citenamefont {Blazit}, \citenamefont
  {Li}, \citenamefont {Gloter} \emph {et~al.}}]{Tizei2020}%
  \BibitemOpen
  \bibfield  {author} {\bibinfo {author} {\bibfnamefont {L.~H.}\ \bibnamefont
  {Tizei}}, \bibinfo {author} {\bibfnamefont {V.}~\bibnamefont {Mkhitaryan}},
  \bibinfo {author} {\bibfnamefont {H.}~\bibnamefont {Louren{\c{c}}o-Martins}},
  \bibinfo {author} {\bibfnamefont {L.}~\bibnamefont {Scarabelli}}, \bibinfo
  {author} {\bibfnamefont {K.}~\bibnamefont {Watanabe}}, \bibinfo {author}
  {\bibfnamefont {T.}~\bibnamefont {Taniguchi}}, \bibinfo {author}
  {\bibfnamefont {M.}~\bibnamefont {Tenc{\'e}}}, \bibinfo {author}
  {\bibfnamefont {J.-D.}\ \bibnamefont {Blazit}}, \bibinfo {author}
  {\bibfnamefont {X.}~\bibnamefont {Li}}, \bibinfo {author} {\bibfnamefont
  {A.}~\bibnamefont {Gloter}}, \emph {et~al.},\ }\bibfield  {title} {\bibinfo
  {title} {Tailored nanoscale plasmon-enhanced vibrational electron
  spectroscopy},\ }\href@noop {} {\bibfield  {journal} {\bibinfo  {journal}
  {Nano Lett.}\ }\textbf {\bibinfo {volume} {20}},\ \bibinfo {pages} {2973}
  (\bibinfo {year} {2020})}\BibitemShut {NoStop}%
\bibitem [{\citenamefont {Krivanek}\ \emph {et~al.}(2014)\citenamefont
  {Krivanek}, \citenamefont {Lovejoy}, \citenamefont {Dellby}, \citenamefont
  {Aoki}, \citenamefont {Carpenter}, \citenamefont {Rez}, \citenamefont
  {Soignard}, \citenamefont {Zhu}, \citenamefont {Batson}, \citenamefont
  {Lagos}, \citenamefont {Egerton},\ and\ \citenamefont {Crozier}}]{KLD14}%
  \BibitemOpen
  \bibfield  {author} {\bibinfo {author} {\bibfnamefont {O.~L.}\ \bibnamefont
  {Krivanek}}, \bibinfo {author} {\bibfnamefont {T.~C.}\ \bibnamefont
  {Lovejoy}}, \bibinfo {author} {\bibfnamefont {N.}~\bibnamefont {Dellby}},
  \bibinfo {author} {\bibfnamefont {T.}~\bibnamefont {Aoki}}, \bibinfo {author}
  {\bibfnamefont {R.~W.}\ \bibnamefont {Carpenter}}, \bibinfo {author}
  {\bibfnamefont {P.}~\bibnamefont {Rez}}, \bibinfo {author} {\bibfnamefont
  {E.}~\bibnamefont {Soignard}}, \bibinfo {author} {\bibfnamefont
  {J.}~\bibnamefont {Zhu}}, \bibinfo {author} {\bibfnamefont {P.~E.}\
  \bibnamefont {Batson}}, \bibinfo {author} {\bibfnamefont {M.~J.}\
  \bibnamefont {Lagos}}, \bibinfo {author} {\bibfnamefont {R.~F.}\ \bibnamefont
  {Egerton}},\ and\ \bibinfo {author} {\bibfnamefont {P.~A.}\ \bibnamefont
  {Crozier}},\ }\bibfield  {title} {\bibinfo {title} {Vibrational spectroscopy
  in the electron microscope},\ }\href@noop {} {\bibfield  {journal} {\bibinfo
  {journal} {Nature}\ }\textbf {\bibinfo {volume} {514}},\ \bibinfo {pages}
  {209} (\bibinfo {year} {2014})}\BibitemShut {NoStop}%
\bibitem [{\citenamefont {Lagos}\ \emph {et~al.}(2017)\citenamefont {Lagos},
  \citenamefont {Tr\"ugler}, \citenamefont {Hohenester},\ and\ \citenamefont
  {Batson}}]{LTHB17}%
  \BibitemOpen
  \bibfield  {author} {\bibinfo {author} {\bibfnamefont {M.~J.}\ \bibnamefont
  {Lagos}}, \bibinfo {author} {\bibfnamefont {A.}~\bibnamefont {Tr\"ugler}},
  \bibinfo {author} {\bibfnamefont {U.}~\bibnamefont {Hohenester}},\ and\
  \bibinfo {author} {\bibfnamefont {P.~E.}\ \bibnamefont {Batson}},\ }\bibfield
   {title} {\bibinfo {title} {Mapping vibrational surface and bulk modes in a
  single nanocube},\ }\href@noop {} {\bibfield  {journal} {\bibinfo  {journal}
  {Nature}\ }\textbf {\bibinfo {volume} {543}},\ \bibinfo {pages} {529}
  (\bibinfo {year} {2017})}\BibitemShut {NoStop}%
\bibitem [{\citenamefont {Qi}\ \emph {et~al.}(2019)\citenamefont {Qi},
  \citenamefont {Wang}, \citenamefont {Li}, \citenamefont {Sun}, \citenamefont
  {Chen}, \citenamefont {Han}, \citenamefont {Li}, \citenamefont {Zhang},
  \citenamefont {Liu}, \citenamefont {Yu} \emph {et~al.}}]{Qi2019}%
  \BibitemOpen
  \bibfield  {author} {\bibinfo {author} {\bibfnamefont {R.}~\bibnamefont
  {Qi}}, \bibinfo {author} {\bibfnamefont {R.}~\bibnamefont {Wang}}, \bibinfo
  {author} {\bibfnamefont {Y.}~\bibnamefont {Li}}, \bibinfo {author}
  {\bibfnamefont {Y.}~\bibnamefont {Sun}}, \bibinfo {author} {\bibfnamefont
  {S.}~\bibnamefont {Chen}}, \bibinfo {author} {\bibfnamefont {B.}~\bibnamefont
  {Han}}, \bibinfo {author} {\bibfnamefont {N.}~\bibnamefont {Li}}, \bibinfo
  {author} {\bibfnamefont {Q.}~\bibnamefont {Zhang}}, \bibinfo {author}
  {\bibfnamefont {X.}~\bibnamefont {Liu}}, \bibinfo {author} {\bibfnamefont
  {D.}~\bibnamefont {Yu}}, \emph {et~al.},\ }\bibfield  {title} {\bibinfo
  {title} {Probing far-infrared surface phonon polaritons in semiconductor
  nanostructures at nanoscale},\ }\href@noop {} {\bibfield  {journal} {\bibinfo
   {journal} {Nano Lett.}\ }\textbf {\bibinfo {volume} {19}},\ \bibinfo {pages}
  {5070} (\bibinfo {year} {2019})}\BibitemShut {NoStop}%
\bibitem [{\citenamefont {Hachtel}\ \emph {et~al.}(2019)\citenamefont
  {Hachtel}, \citenamefont {Huang}, \citenamefont {Popovs}, \citenamefont
  {Jansone-Popova}, \citenamefont {Keum}, \citenamefont {Jakowski},
  \citenamefont {Lovejoy}, \citenamefont {Dellby}, \citenamefont {Krivanek},\
  and\ \citenamefont {Idrobo}}]{Hachtel2019}%
  \BibitemOpen
  \bibfield  {author} {\bibinfo {author} {\bibfnamefont {J.~A.}\ \bibnamefont
  {Hachtel}}, \bibinfo {author} {\bibfnamefont {J.}~\bibnamefont {Huang}},
  \bibinfo {author} {\bibfnamefont {I.}~\bibnamefont {Popovs}}, \bibinfo
  {author} {\bibfnamefont {S.}~\bibnamefont {Jansone-Popova}}, \bibinfo
  {author} {\bibfnamefont {J.~K.}\ \bibnamefont {Keum}}, \bibinfo {author}
  {\bibfnamefont {J.}~\bibnamefont {Jakowski}}, \bibinfo {author}
  {\bibfnamefont {T.~C.}\ \bibnamefont {Lovejoy}}, \bibinfo {author}
  {\bibfnamefont {N.}~\bibnamefont {Dellby}}, \bibinfo {author} {\bibfnamefont
  {O.~L.}\ \bibnamefont {Krivanek}},\ and\ \bibinfo {author} {\bibfnamefont
  {J.~C.}\ \bibnamefont {Idrobo}},\ }\bibfield  {title} {\bibinfo {title}
  {Identification of site-specific isotopic labels by vibrational spectroscopy
  in the electron microscope},\ }\href@noop {} {\bibfield  {journal} {\bibinfo
  {journal} {Science}\ }\textbf {\bibinfo {volume} {363}},\ \bibinfo {pages}
  {525} (\bibinfo {year} {2019})}\BibitemShut {NoStop}%
\bibitem [{\citenamefont {Hage}\ \emph {et~al.}(2020)\citenamefont {Hage},
  \citenamefont {Radtke}, \citenamefont {Kepaptsoglou}, \citenamefont
  {Lazzeri},\ and\ \citenamefont {Ramasse}}]{Hage2020}%
  \BibitemOpen
  \bibfield  {author} {\bibinfo {author} {\bibfnamefont {F.}~\bibnamefont
  {Hage}}, \bibinfo {author} {\bibfnamefont {G.}~\bibnamefont {Radtke}},
  \bibinfo {author} {\bibfnamefont {D.}~\bibnamefont {Kepaptsoglou}}, \bibinfo
  {author} {\bibfnamefont {M.}~\bibnamefont {Lazzeri}},\ and\ \bibinfo {author}
  {\bibfnamefont {Q.}~\bibnamefont {Ramasse}},\ }\bibfield  {title} {\bibinfo
  {title} {Single-atom vibrational spectroscopy in the scanning transmission
  electron microscope},\ }\href@noop {} {\bibfield  {journal} {\bibinfo
  {journal} {Science}\ }\textbf {\bibinfo {volume} {367}},\ \bibinfo {pages}
  {1124} (\bibinfo {year} {2020})}\BibitemShut {NoStop}%
\bibitem [{\citenamefont {Losquin}\ \emph {et~al.}(2015)\citenamefont
  {Losquin}, \citenamefont {Zagonel}, \citenamefont {Myroshnychenko},
  \citenamefont {Rodr\'{\i}guez-Gonz\'alez}, \citenamefont {Tenc\'e},
  \citenamefont {Scarabelli}, \citenamefont {F\"orstner}, \citenamefont
  {Liz-Marz\'an}, \citenamefont {de~Abajo}, \citenamefont {St\'ephan},\ and\
  \citenamefont {Kociak}}]{paper251}%
  \BibitemOpen
  \bibfield  {author} {\bibinfo {author} {\bibfnamefont {A.}~\bibnamefont
  {Losquin}}, \bibinfo {author} {\bibfnamefont {L.~F.}\ \bibnamefont
  {Zagonel}}, \bibinfo {author} {\bibfnamefont {V.}~\bibnamefont
  {Myroshnychenko}}, \bibinfo {author} {\bibfnamefont {B.}~\bibnamefont
  {Rodr\'{\i}guez-Gonz\'alez}}, \bibinfo {author} {\bibfnamefont
  {M.}~\bibnamefont {Tenc\'e}}, \bibinfo {author} {\bibfnamefont
  {L.}~\bibnamefont {Scarabelli}}, \bibinfo {author} {\bibfnamefont
  {J.}~\bibnamefont {F\"orstner}}, \bibinfo {author} {\bibfnamefont {L.~M.}\
  \bibnamefont {Liz-Marz\'an}}, \bibinfo {author} {\bibfnamefont {F.~J.~G.}\
  \bibnamefont {de~Abajo}}, \bibinfo {author} {\bibfnamefont {O.}~\bibnamefont
  {St\'ephan}},\ and\ \bibinfo {author} {\bibfnamefont {M.}~\bibnamefont
  {Kociak}},\ }\bibfield  {title} {\bibinfo {title} {Unveiling nanometer scale
  extinction and scattering phenomena through combined electron energy loss
  spectroscopy and cathodoluminescence measurements},\ }\href@noop {}
  {\bibfield  {journal} {\bibinfo  {journal} {Nano\ Lett.}\ }\textbf {\bibinfo
  {volume} {15}},\ \bibinfo {pages} {1229} (\bibinfo {year}
  {2015})}\BibitemShut {NoStop}%
\bibitem [{\citenamefont {Couillard}\ \emph {et~al.}(2011)\citenamefont
  {Couillard}, \citenamefont {Radtke}, \citenamefont {Knights},\ and\
  \citenamefont {Botton}}]{Couillard2011}%
  \BibitemOpen
  \bibfield  {author} {\bibinfo {author} {\bibfnamefont {M.}~\bibnamefont
  {Couillard}}, \bibinfo {author} {\bibfnamefont {G.}~\bibnamefont {Radtke}},
  \bibinfo {author} {\bibfnamefont {A.~P.}\ \bibnamefont {Knights}},\ and\
  \bibinfo {author} {\bibfnamefont {G.~A.}\ \bibnamefont {Botton}},\ }\bibfield
   {title} {\bibinfo {title} {Three-dimensional atomic structure of metastable
  nanoclusters in doped semiconductors},\ }\href@noop {} {\bibfield  {journal}
  {\bibinfo  {journal} {Phys. Rev. Lett.}\ }\textbf {\bibinfo {volume} {107}},\
  \bibinfo {pages} {186104} (\bibinfo {year} {2011})}\BibitemShut {NoStop}%
\bibitem [{\citenamefont {Johnstone}\ \emph {et~al.}(2020)\citenamefont
  {Johnstone}, \citenamefont {Crout}, \citenamefont {Nord}, \citenamefont
  {Laulainen}, \citenamefont {Høgås}, \citenamefont {EirikOpheim},
  \citenamefont {Martineau}, \citenamefont {Bergh}, \citenamefont {Francis},
  \citenamefont {Smeets}, \citenamefont {Prestat}, \citenamefont {andrew
  ross1}, \citenamefont {Collins}, \citenamefont {Hjorth}, \citenamefont
  {Mohsen}, \citenamefont {Furnival}, \citenamefont {Jannis}, \citenamefont
  {Jacobsen}, \citenamefont {AndrewHerzing}, \citenamefont {Poon},
  \citenamefont {Ånes}, \citenamefont {Morzy}, \citenamefont {phillipcrout},
  \citenamefont {Doherty}, \citenamefont {affaniqbal}, \citenamefont
  {Ostasevicius}, \citenamefont {mvonlany},\ and\ \citenamefont
  {Tovey}}]{duncan_n_johnstone_2020_4302056}%
  \BibitemOpen
  \bibfield  {author} {\bibinfo {author} {\bibfnamefont {D.~N.}\ \bibnamefont
  {Johnstone}}, \bibinfo {author} {\bibfnamefont {P.}~\bibnamefont {Crout}},
  \bibinfo {author} {\bibfnamefont {M.}~\bibnamefont {Nord}}, \bibinfo {author}
  {\bibfnamefont {J.}~\bibnamefont {Laulainen}}, \bibinfo {author}
  {\bibfnamefont {S.}~\bibnamefont {Høgås}}, \bibinfo {author} {\bibnamefont
  {EirikOpheim}}, \bibinfo {author} {\bibfnamefont {B.}~\bibnamefont
  {Martineau}}, \bibinfo {author} {\bibfnamefont {T.}~\bibnamefont {Bergh}},
  \bibinfo {author} {\bibfnamefont {C.}~\bibnamefont {Francis}}, \bibinfo
  {author} {\bibfnamefont {S.}~\bibnamefont {Smeets}}, \bibinfo {author}
  {\bibfnamefont {E.}~\bibnamefont {Prestat}}, \bibinfo {author} {\bibnamefont
  {andrew ross1}}, \bibinfo {author} {\bibfnamefont {S.}~\bibnamefont
  {Collins}}, \bibinfo {author} {\bibfnamefont {I.}~\bibnamefont {Hjorth}},
  \bibinfo {author} {\bibnamefont {Mohsen}}, \bibinfo {author} {\bibfnamefont
  {T.}~\bibnamefont {Furnival}}, \bibinfo {author} {\bibfnamefont
  {D.}~\bibnamefont {Jannis}}, \bibinfo {author} {\bibfnamefont
  {E.}~\bibnamefont {Jacobsen}}, \bibinfo {author} {\bibnamefont
  {AndrewHerzing}}, \bibinfo {author} {\bibfnamefont {T.}~\bibnamefont {Poon}},
  \bibinfo {author} {\bibfnamefont {H.~W.}\ \bibnamefont {Ånes}}, \bibinfo
  {author} {\bibfnamefont {J.}~\bibnamefont {Morzy}}, \bibinfo {author}
  {\bibnamefont {phillipcrout}}, \bibinfo {author} {\bibfnamefont
  {T.}~\bibnamefont {Doherty}}, \bibinfo {author} {\bibnamefont {affaniqbal}},
  \bibinfo {author} {\bibfnamefont {T.}~\bibnamefont {Ostasevicius}}, \bibinfo
  {author} {\bibnamefont {mvonlany}},\ and\ \bibinfo {author} {\bibfnamefont
  {R.}~\bibnamefont {Tovey}},\ }\href {https://doi.org/10.5281/zenodo.4302056}
  {\bibinfo {title} {pyxem/pyxem: pyxem 0.12.3}} (\bibinfo {year}
  {2020})\BibitemShut {NoStop}%
\bibitem [{\citenamefont {de~la Peña}\ \emph {et~al.}(2020)\citenamefont
  {de~la Peña}, \citenamefont {Prestat}, \citenamefont {Fauske}, \citenamefont
  {Burdet}, \citenamefont {Furnival}, \citenamefont {Jokubauskas},
  \citenamefont {Nord}, \citenamefont {Ostasevicius}, \citenamefont
  {MacArthur}, \citenamefont {Johnstone}, \citenamefont {Sarahan},
  \citenamefont {Lähnemann}, \citenamefont {Taillon}, \citenamefont {pquinn
  dls}, \citenamefont {Aarholt}, \citenamefont {Migunov}, \citenamefont
  {Eljarrat}, \citenamefont {Caron}, \citenamefont {Mazzucco}, \citenamefont
  {Martineau}, \citenamefont {Somnath}, \citenamefont {Poon}, \citenamefont
  {Walls}, \citenamefont {Slater}, \citenamefont {actions user}, \citenamefont
  {Tappy}, \citenamefont {Cautaerts}, \citenamefont {Winkler}, \citenamefont
  {Donval},\ and\ \citenamefont {Myers}}]{francisco_de_la_pena_2020_4294676}%
  \BibitemOpen
  \bibfield  {author} {\bibinfo {author} {\bibfnamefont {F.}~\bibnamefont
  {de~la Peña}}, \bibinfo {author} {\bibfnamefont {E.}~\bibnamefont
  {Prestat}}, \bibinfo {author} {\bibfnamefont {V.~T.}\ \bibnamefont {Fauske}},
  \bibinfo {author} {\bibfnamefont {P.}~\bibnamefont {Burdet}}, \bibinfo
  {author} {\bibfnamefont {T.}~\bibnamefont {Furnival}}, \bibinfo {author}
  {\bibfnamefont {P.}~\bibnamefont {Jokubauskas}}, \bibinfo {author}
  {\bibfnamefont {M.}~\bibnamefont {Nord}}, \bibinfo {author} {\bibfnamefont
  {T.}~\bibnamefont {Ostasevicius}}, \bibinfo {author} {\bibfnamefont {K.~E.}\
  \bibnamefont {MacArthur}}, \bibinfo {author} {\bibfnamefont {D.~N.}\
  \bibnamefont {Johnstone}}, \bibinfo {author} {\bibfnamefont {M.}~\bibnamefont
  {Sarahan}}, \bibinfo {author} {\bibfnamefont {J.}~\bibnamefont {Lähnemann}},
  \bibinfo {author} {\bibfnamefont {J.}~\bibnamefont {Taillon}}, \bibinfo
  {author} {\bibnamefont {pquinn dls}}, \bibinfo {author} {\bibfnamefont
  {T.}~\bibnamefont {Aarholt}}, \bibinfo {author} {\bibfnamefont
  {V.}~\bibnamefont {Migunov}}, \bibinfo {author} {\bibfnamefont
  {A.}~\bibnamefont {Eljarrat}}, \bibinfo {author} {\bibfnamefont
  {J.}~\bibnamefont {Caron}}, \bibinfo {author} {\bibfnamefont
  {S.}~\bibnamefont {Mazzucco}}, \bibinfo {author} {\bibfnamefont
  {B.}~\bibnamefont {Martineau}}, \bibinfo {author} {\bibfnamefont
  {S.}~\bibnamefont {Somnath}}, \bibinfo {author} {\bibfnamefont
  {T.}~\bibnamefont {Poon}}, \bibinfo {author} {\bibfnamefont {M.}~\bibnamefont
  {Walls}}, \bibinfo {author} {\bibfnamefont {T.}~\bibnamefont {Slater}},
  \bibinfo {author} {\bibnamefont {actions user}}, \bibinfo {author}
  {\bibfnamefont {N.}~\bibnamefont {Tappy}}, \bibinfo {author} {\bibfnamefont
  {N.}~\bibnamefont {Cautaerts}}, \bibinfo {author} {\bibfnamefont
  {F.}~\bibnamefont {Winkler}}, \bibinfo {author} {\bibfnamefont
  {G.}~\bibnamefont {Donval}},\ and\ \bibinfo {author} {\bibfnamefont {J.~C.}\
  \bibnamefont {Myers}},\ }\href {https://doi.org/10.5281/zenodo.4294676}
  {\bibinfo {title} {hyperspy/hyperspy: Release v1.6.1}} (\bibinfo {year}
  {2020})\BibitemShut {NoStop}%
\bibitem [{\citenamefont {Pizzocchero}\ \emph {et~al.}(2016)\citenamefont
  {Pizzocchero}, \citenamefont {Gammelgaard}, \citenamefont {Jessen},
  \citenamefont {Caridad}, \citenamefont {Wang}, \citenamefont {Hone},
  \citenamefont {B{\o}ggild},\ and\ \citenamefont {Booth}}]{Pizzocchero2016}%
  \BibitemOpen
  \bibfield  {author} {\bibinfo {author} {\bibfnamefont {F.}~\bibnamefont
  {Pizzocchero}}, \bibinfo {author} {\bibfnamefont {L.}~\bibnamefont
  {Gammelgaard}}, \bibinfo {author} {\bibfnamefont {B.~S.}\ \bibnamefont
  {Jessen}}, \bibinfo {author} {\bibfnamefont {J.~M.}\ \bibnamefont {Caridad}},
  \bibinfo {author} {\bibfnamefont {L.}~\bibnamefont {Wang}}, \bibinfo {author}
  {\bibfnamefont {J.}~\bibnamefont {Hone}}, \bibinfo {author} {\bibfnamefont
  {P.}~\bibnamefont {B{\o}ggild}},\ and\ \bibinfo {author} {\bibfnamefont
  {T.~J.}\ \bibnamefont {Booth}},\ }\bibfield  {title} {\bibinfo {title} {The
  hot pick-up technique for batch assembly of van der waals heterostructures},\
  }\href@noop {} {\bibfield  {journal} {\bibinfo  {journal} {Nature Com.}\
  }\textbf {\bibinfo {volume} {7}},\ \bibinfo {pages} {1} (\bibinfo {year}
  {2016})}\BibitemShut {NoStop}%
\bibitem [{\citenamefont {Taniguchi}\ and\ \citenamefont
  {Watanabe}(2007)}]{Taniguchi2007}%
  \BibitemOpen
  \bibfield  {author} {\bibinfo {author} {\bibfnamefont {T.}~\bibnamefont
  {Taniguchi}}\ and\ \bibinfo {author} {\bibfnamefont {K.}~\bibnamefont
  {Watanabe}},\ }\bibfield  {title} {\bibinfo {title} {Synthesis of high-purity
  boron nitride single crystals under high pressure by using ba-bn solvent},\
  }\href@noop {} {\bibfield  {journal} {\bibinfo  {journal} {Journal of Crystal
  Growth}\ }\textbf {\bibinfo {volume} {303}},\ \bibinfo {pages} {525}
  (\bibinfo {year} {2007})}\BibitemShut {NoStop}%
\bibitem [{\citenamefont {St\'ephan}\ \emph {et~al.}(2002)\citenamefont
  {St\'ephan}, \citenamefont {Taverna}, \citenamefont {Kociak}, \citenamefont
  {Suenaga}, \citenamefont {Henrard},\ and\ \citenamefont
  {Colliex}}]{Stephan2002}%
  \BibitemOpen
  \bibfield  {author} {\bibinfo {author} {\bibfnamefont {O.}~\bibnamefont
  {St\'ephan}}, \bibinfo {author} {\bibfnamefont {D.}~\bibnamefont {Taverna}},
  \bibinfo {author} {\bibfnamefont {M.}~\bibnamefont {Kociak}}, \bibinfo
  {author} {\bibfnamefont {K.}~\bibnamefont {Suenaga}}, \bibinfo {author}
  {\bibfnamefont {L.}~\bibnamefont {Henrard}},\ and\ \bibinfo {author}
  {\bibfnamefont {C.}~\bibnamefont {Colliex}},\ }\bibfield  {title} {\bibinfo
  {title} {Dielectric response of isolated carbon nanotubes investigated by
  spatially resolved electron energy-loss spectroscopy: From multiwalled to
  single-walled nanotubes},\ }\href
  {https://doi.org/10.1103/PhysRevB.66.155422} {\bibfield  {journal} {\bibinfo
  {journal} {Phys. Rev. B}\ }\textbf {\bibinfo {volume} {66}},\ \bibinfo
  {pages} {155422} (\bibinfo {year} {2002})}\BibitemShut {NoStop}%
\end{thebibliography}%

	\subsection{Acknowledgments}
	This project has been funded in part by the National Agency for Research under the program of future investment TEMPOS-CHROMATEM (reference no. ANR-10-EQPX-50) and from the European Union’s Horizon 2020 research and innovation programme under grant agreement No 823717 (ESTEEM3) and 101017720 (EBEAM). K.W. and T.T. acknowledge support from the Elemental Strategy Initiative conducted by the MEXT, Japan, Grant Number JPMXP0112101001, JSPS KAKENHI Grant Number JP20H00354 and the CREST(JPMJCR15F3), JST. This work has been supported by Region \^Ile-de-France in the framework of DIM SIRTEQ". We thank NION, HennyZ, and Attolight for the helpfull interaction on the adaptation of the CL system to the NION sample chamber on the ChromaTEM microscope and customization of the liquid nitrogen sample holder. We acknowledge the joint effort of the STEM team at the LPS-Orsay and, in particular Marcel Tenc\'e and Xiaoyan Li, concerning instrumental developments. We thank Ashish Arora and co-authors for kindly providing the optical absorption data on a WS$_2$ monolayer encapsulated in h-BN. Luiz F. Zagonel is acknowledged for ideas and discussion on data analysis. 
	
	\subsection{Competing interests} MK patented and licensed technologies at the basis of the Attolight M\"onch used in this study, and is a part time consultant at Attolight.  All other authors declare no competing financial interests.
	
	\newpage
	
	\renewcommand\thefigure{SI\arabic{figure}} 
	\setcounter{figure}{0} 
	\section{Supplementary information to Nanoscale modification of WS$_2$ trion emission by its local electromagnetic environment}
	
	\subsection{Methods}

	Scanning transmission electron microscopy (STEM) imaging, diffraction, CL and EELS experiments were performed on a modified Nion Hermes200 operated at 60 and 100 keV. In this microscope, subnanometer electron beams with sub 10 meV energy spread \cite{Tizei2020} can be generated for high spatial resolution imaging, diffraction and spectroscopy. High energy and spatial resolution have been substantially improved over the last ten years due to new monochromator technologies \cite{KLD14, Tizei2015, LTHB17, Qi2019,Hachtel2019, Hage2020}. The energy resolution of the EELS data presented here was between 20 and 30 meV (energy width of the primary electron beam). CL used a Mönch system from Attolight \cite{Kociak2017}, with an energy resolution of 8 meV (minimum separation between two discernible emission peaks). Combined EELS-CL experiments have been used in the past to understand optical extinction and scattering in metallic plasmonic nanoparticles \cite{paper251}, but required much smaller requirements on the spectral resolution, due to the very large linewidth of the plasmons. Sample were kept at 150 K using a liquid nitrogen HennyZ sample holder for spectroscopic experiments, except for EELS chemical mapping. The typical exposure time used for CL and EELS low-loss experiment were 300 ms, and  for core-loss and diffraction, 50 ms.	
	
\begin{figure} [H]
	\centering
		\includegraphics[width=12cm]{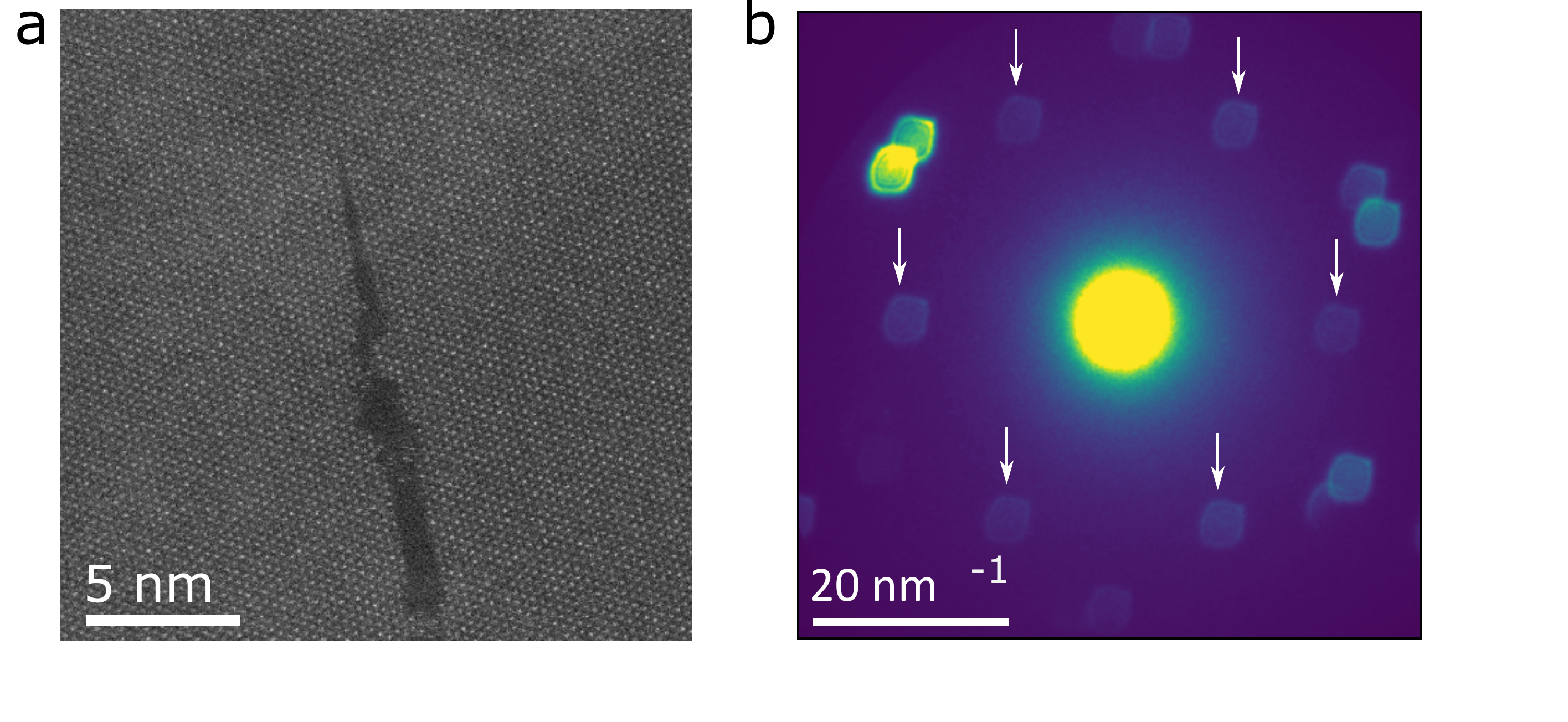}
		\caption{\textbf{Atomically resolved HAADF image and diffraction of a WS$_2$ monolayer:   } \textbf{(a)} Atomically resolved image of a WS$_2$ monolayer encapsulated in h-BN. The h-BN layer is barely visible due to off-axis imaging. \textbf{(b)} Diffraction pattern of the same sample showing the monolayer, a faint hexagon pattern of the first-order reflections marked by white arrows. The extra spots come from diffraction on the two h-BN crystals.}
		\label{SI_AtomicImage_Diffraction}
	\end{figure}

	Atomically resolved imaging and spatially resolved EELS chemical maps were acquired on Nion UltraSTEM 200 operated at 100 keV, with the samples at room temperature. All images shown are high angle annular dark field (HAADF) images, in which the intensity is proportional to the projected atomic number, with W atoms showing as bright dots. The columns with two S atoms in projection are harder to pinpoint due to the background created by scattering in the h-BN layers. Diffraction effects play a smaller role in HAADF image intensity, so imaging with the h-BN slightly off-axis is beneficial to observe the single WS$_2$ monolayer embedded in 25 nm of h-BN, as demonstrated before for CeSi clusters in Si matrices \cite{Couillard2011}. Encapsulation has also ensured a high stability of the monolayers under 60 keV and 100 keV electron irradiation, allowing imaging of free edges (Fig. \ref{Figure_Experiment}a and Fig. \ref{SI_AtomicImage_Diffraction}).
	
	Diffraction patterns were acquired for each beam position on the sample with typical convergence semi-angle between 3 and 5 mrad and the camera dwell time was 50 ms. Diffraction mapping analysis was done with the Pyxem \cite{duncan_n_johnstone_2020_4302056} and Hyperspy \cite{francisco_de_la_pena_2020_4294676} python libraries. The quantity calculated is the displacement gradient tensor, corresponding to the difference between the deformed vectors and a reference, where the reference is a calculated vector with the reciprocal space length and orientation for unstrained WS$_2$. A right-handed polar decomposition was used to separate deformation and rotation. The angle of rotation was then recovered from the rotation matrix.
	Each deformed vector was defined with the barycenter of the diffraction spots, and the center of the direct beam was first aligned with a cross correlation method. The barycenter method is limited by illumination changes of the diffraction spots (due to diffraction on the thicker h-BN layers), limiting our current strain measurements to above 1$\%$.

	The CL and EELS datacube spectral fitting was done with the Hyperspy bounded multifit tool, using gaussian profiles. The Lorentzian profile was also tried since the EELS exciton peak profiles should be close to lorentzian, but the results are not displayed here to keep coherence for all curve fits.
	Most of the values found in the text are from mean spectra extracted from different regions of interest in each spectrum-image (these extracted spectra have a much higher signal to noise ratio). The extracted spectra are fitted with gaussian profiles, and the uncertainty associated is the standard deviation of the fit.
	
	\subsubsection{BSS+PCA analysis description} The blind source separation (BSS) technique consists of the separation of a mixed signal into individual components. The algorithm used in this paper was the independent component analysis (ICA) implemented in Hyperspy. In the ICA algorithm, the individual components are additive, and treated as non-gaussian and statistically independent. We chose 3 components, the first one contains the background, the h-BN and the carbon that are correlated together. This carbon can be from contamination of the sample. The second one contains the silicon and some carbon, and the third one contains mostly noise.

	\subsubsection{Sample preparation for electron spectroscopy and microscopy}
	
	h-BN/WS$_2$/h-BN heterostructures were fabricated by using modified dry transfer method \cite{Pizzocchero2016} then transferred to a TEM grid. WS$_2$ was purchased from 2DSemiconductors and high quality h-BN synthesized by high pressure-high temperature method \cite{Taniguchi2007} was used. All constituted layers were first exfoliated onto a SiO$_2$/Si substrate using the scotch-tape method \cite{Mak2010}. A PDMS (polydimethylsiloxane) mask spin-coated by 15 $\%$ PPC (polypropylene carbonate) is used for polymer stamp. PDMS is made by using the 20:1 ratio of Sylgard 184 pro-polymer to curing agent and kept at ambient conditions for overnight. To enhance adhesion between PDMS and PPC, PDMS mask was treated by oxygen plasma (18 W) for 5 min, before the spin-coating of PPC at 3000 rpm followed by heat treatment at 160 $^\circ$C for 10 minutes. This polymer stamp was mounted to the micromanipulator upside down. A brief description of the procedures are described below.
	\begin{enumerate}
	  \item Pick-up exfoliated h-BN crystal from SiO$_2$/Si substrate by contacting polymer stamp to the target crystal at 50 $^\circ$C for 1 minute and lifting the stamp.
	  \item Repeat step 1 to pick-up the monolayer WS$_2$ crystal and bottom h-BN layer subsequently.
	  \item Drop-down the stack (h-BN/WS$_2$/h-BN) on a new SiO$_2$/Si substrate by contact at higher temperature (120 $^\circ$C)  for 10 minutes.
	  \item Clean the PPC residue on the surface with acetone and IPA.
	  \item Anneal the heterostructure to enhance the interlayer coupling between constituent layers at 250 120 $^\circ$C)  for 6 hours in Ar environment.
	  \item Spin-coat with polymethylmethacrylate (PMMA, 495K, Microchem) over the heterostructure at 3000 rpm followed by a heat treatment at 180 $^\circ$C for 5 minutes.
	  \item Etch SiO$_2$/Si substrate by immerse the sample into KOH solution (1M) overnight.
	  \item Transfer the sample to TEM grid (C-flat holey carbon grid with 2 $\mu$m hole diameter) 
	  \item Remove PMMA residue by cleaning with acetone and IPA.
	\end{enumerate}

	\subsection{EELS and optical absorption comparison}
	
	The electromagnetic response properties of materials are usually described by its dielectric function $\epsilon(\omega) = \epsilon_1 (\omega) + i*\epsilon_2(\omega)$. However optical measurements, usually give access to the complex refractive index, $\tilde{n}(\omega)  = n(\omega)+i*\kappa(\omega)$. These two quantities are linked by the relation $\tilde{n}^2(\omega) = \epsilon(\omega)$, which also links their real and imaginary parts by
	
	\begin{equation}
		\epsilon_1(\omega) = n^2(\omega)-\kappa^2(\omega)
		\epsilon_2(\omega) = 2n(\omega)\kappa(\omega),
	\end{equation}
	
$\kappa$ is the extinction coefficient, which is linked absorption coefficient, $\alpha$, by 
	
	\begin{equation}
		\alpha = \frac{4\pi\kappa}{\lambda},
	\end{equation}
with $\lambda$ the wavelength of light. These two quantities, $\kappa$ and $\alpha$, are linked to the decrease of the total intensity being transmitted through a medium.

	A large part of optical absorption measurements are made from reflectivity, which gives access to reflectance, R:
	
	\begin{equation}
		R = \frac{\tilde{n}-1}{\tilde{n}+1} = \frac{(n-1)^2+\kappa^2}{(n+1)^2+\kappa^2},
	\end{equation}
from which $\epsilon_1(\omega)$ and $\epsilon_2(\omega)$ can be calculated using the Kramers-Kronig relation and a model taking into account the different dielectric layers in the sample under study.
	
	Complications arise from the model necessary to extract the complex dielectric function, which can lead to modifications of line shapes. Therefore, measurements of an absorbance spectrum $ A(\lambda) = 1-R-T$ can be made, from which line shapes can be directly compared. However, this quantity is not a direct measure of any of the materials macroscopic constants. Of course, these can be extracted from the data.
		
\begin{figure}[H]
	\centering
		\includegraphics[width=8cm]{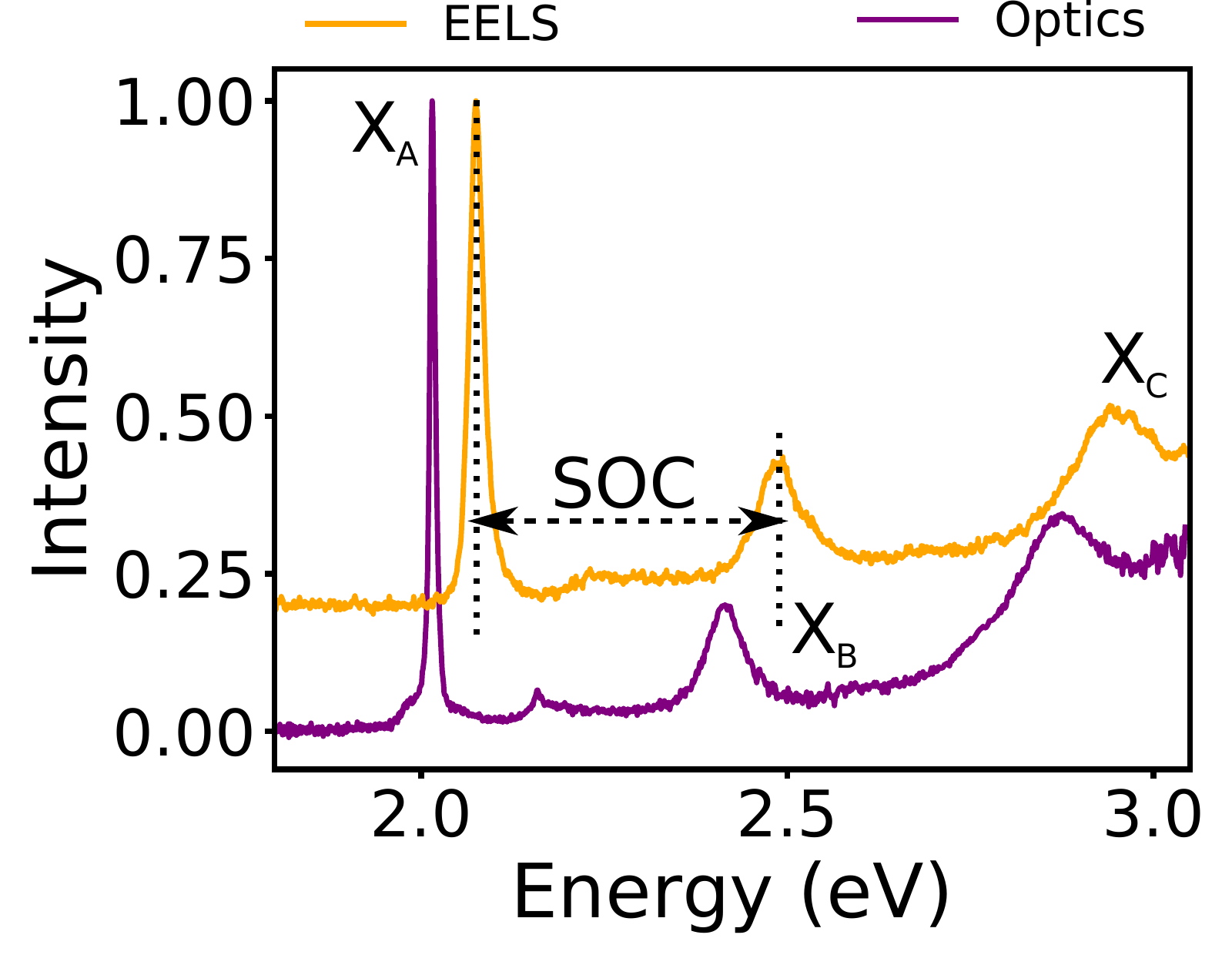}
		\caption{\textbf{EELS and optical absorption comparison:} EELS (orange) spectrum of a WS$_2$ monolayer encapsulated with h-BN. A comparison to optical absorption (purple), from Arora \textit{et al.} \cite{Arora2020}, shows a near perfect one to one correspondence, with the X$_A$, X$_B$ and X$_C$ excitons shown. The extra absorption between X$_A$ and X$_B$ is attributed to the 2s excited state of X$_A$. }
		\label{SI_EELS_Optics}
	\end{figure}

	EELS from an object with dielectric function $\epsilon(\omega)$ measures $Im{\{-1/\epsilon(\omega)\}}$. Therefore, in general, Kramers-Kronig transformation is required to retrieve $\epsilon_1(\omega)$ and $\epsilon_2(\omega)$. However, for atomically thin objects it can be proven that  $Im{\{\epsilon(\omega)\} = \epsilon_2(\omega)}$ is true. In this case, a direct comparison of $\epsilon_2(\omega)$ measured by EELS for atomically thin layers and the calculated value from optical reflectivity is justified. Being a direct measure of $\epsilon_2(\omega)$, one can compare the line shapes of EELS with those in optical absorbance spectra (Fig. \ref{SI_EELS_Optics}). However, a comparison of exact energy positions requires one to take into account the dispersion of the real part of the dielectric function.
	
	As discussed in the text, the energy shifts observed can be due to sample heterogeneity (Fig. \ref{SI_histograms} show that the EELS spectra shift by at least 20 meV within our samples). But part of it can also come from the dispersion of $\epsilon_1(\omega)$. Nevertheless, we do not exclude fine differences between the quantities measured in EELS and optical absorption, in the tens of meV range. The fact that EELS for atomically thin structures measures $Im{\{\epsilon(\omega)\}}$ has only been demonstrated at higher energies, with poorer energy precision (at 15 eV with 300 meV precision for single wall carbon nanotubes \cite{Stephan2002}).
	
		\newpage

\subsection{Supplemental figures}

	\begin{figure}
		\includegraphics[width=12cm]{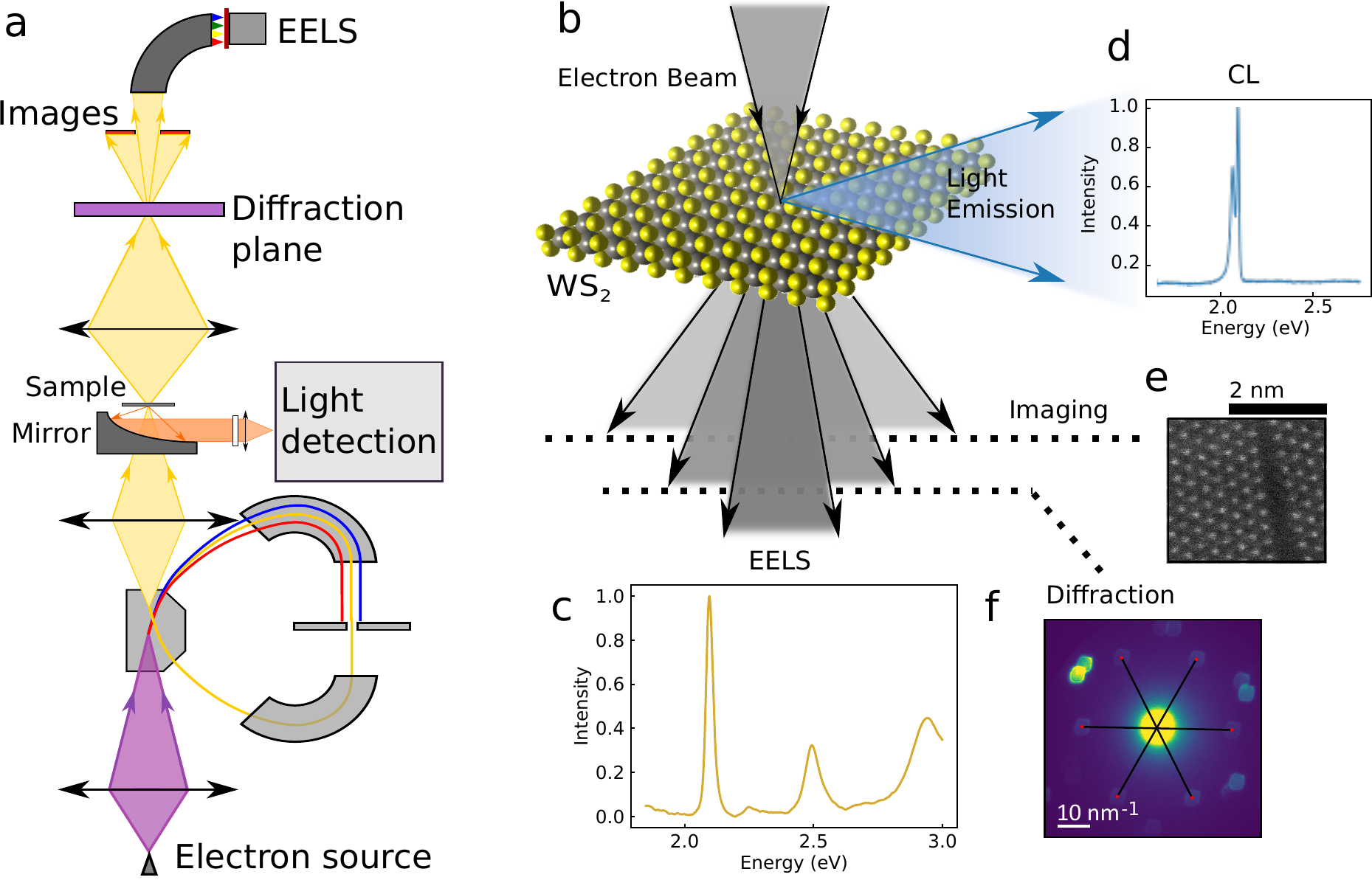}
		\caption{\textbf{Scheme of the experiment:} as described in the main text the microscope used for spectroscopic measurements is equipped with an electron monochromator and light collection system, to allow high resolution EELS and CL. In a scanning transmission electron microscope, a focused electron beam is used. Different signals can be acquired as a function of position, including structural information (diffraction and imaging) and spectroscopic (EELS and CL here). This information can be then correlated, allowing, for example, measurements of the Stokes shift at a given position, or variations of the chemical composition.}
		\label{SI_experiment}
	\end{figure}
	
	\begin{figure}
	\includegraphics[width=12cm]{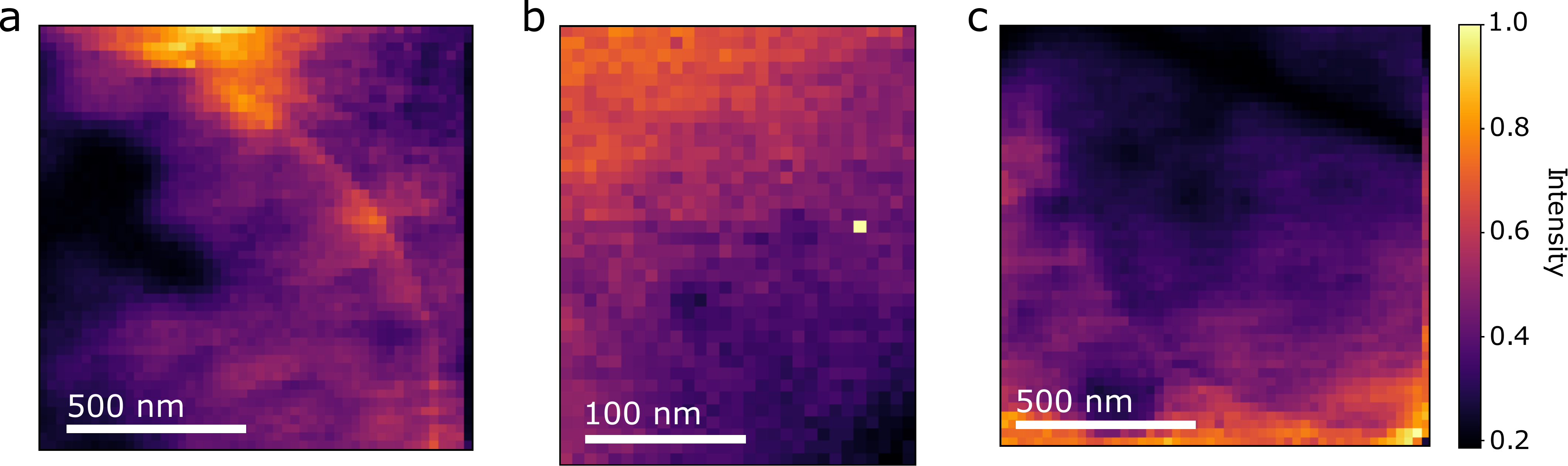}
	\caption{\textbf{X$_A$ intensity maps from main text:} \textbf{(a)} X$_A$ intensity from same area as Fig. \ref{Local_Trion_L1}d, \textbf{(b)} X$_A$ intensity from same area as Fig. \ref{Trion_impurities}b, \textbf{(c)} X$_A$ intensity from same area as Fig. \ref{Trion_interface}b. All maps have been normalized by the maximum of X$_A$ emission in the spectral integrating range.}
	\label{SI_XA_intensity}
	\end{figure}

	\begin{figure}
		\includegraphics[width=16cm]{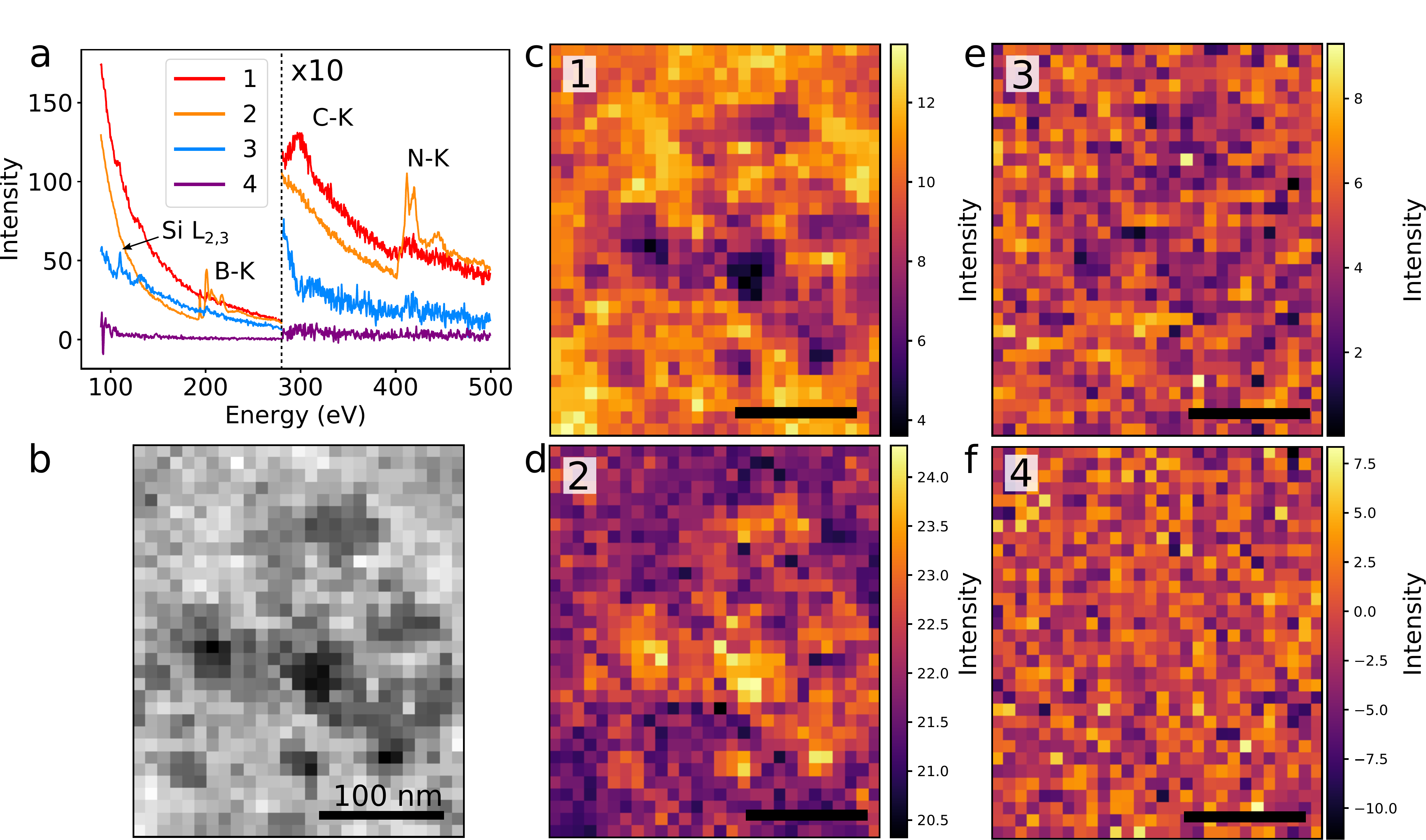}
		\caption{\textbf{Blind source separation (BSS) analysis of Fig. \ref{Trion_impurities}:} \textbf{(a)} BSS componenents spectra, at a normal scale on the left-hand part, and multiplied by 10 for lisibility on the right-hand part, \textbf{(b)} HAADF image of the studied area, \textbf{(c-f)} map of each three components in (a), (c) is 1, (d) is 2, (e) is 3, (f) is 4. The one shown in Fig. \ref{Trion_impurities}d, is component 1 containing some background, oxydized silicon, and carbon that are probably from the PDMS sample preparation.}
		\label{SI_BSS}
	\end{figure}

	\begin{figure}
		\includegraphics[width=16cm]{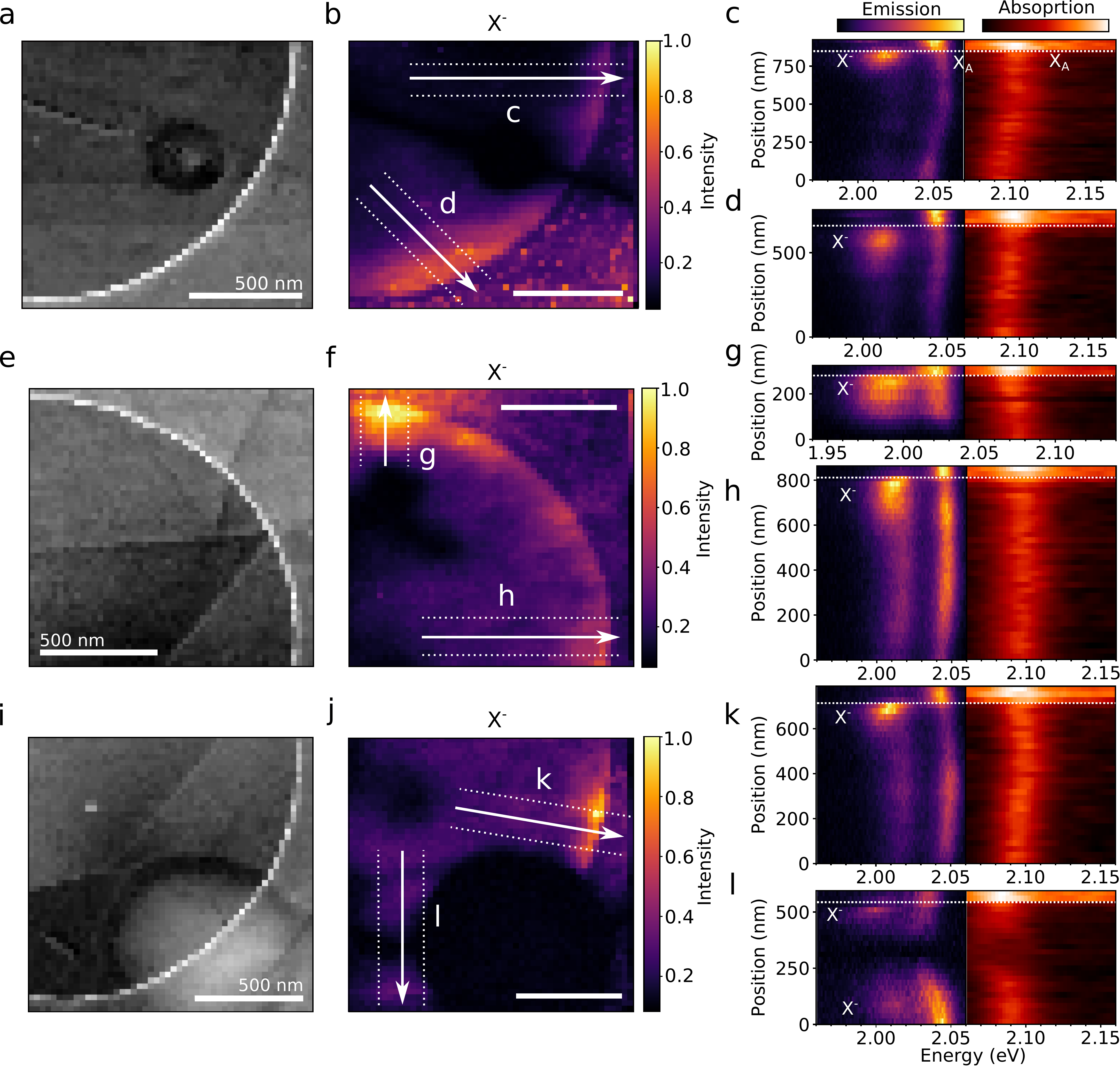}
		\caption{\textbf{Three regions showing similar behavior to that described in Fig. \ref{Trion_interface}b:} \textbf{(a, e, i)} HAADF images of the three regions. \textbf{(b, f, j)} X$^-$ intensity maps similar to that in Fig. \ref{Local_Trion_L1}b, showing an increase in X$^{-}$ emission close to the hole edges. \textbf{(c, d, g, h, k, l)} Spectra from selected profiles in the three regions, showing the change in emission and absorption along the corresponding arrows. Spectra such as (c, h, k) are typically observed next to the carbon membrane support (represented by a dotted line the in the spectra), the others show a less typical behaviour.}
		\label{SI_TrionsOtherHoles}
	\end{figure}

	\begin{figure}
		\includegraphics[width=16cm]{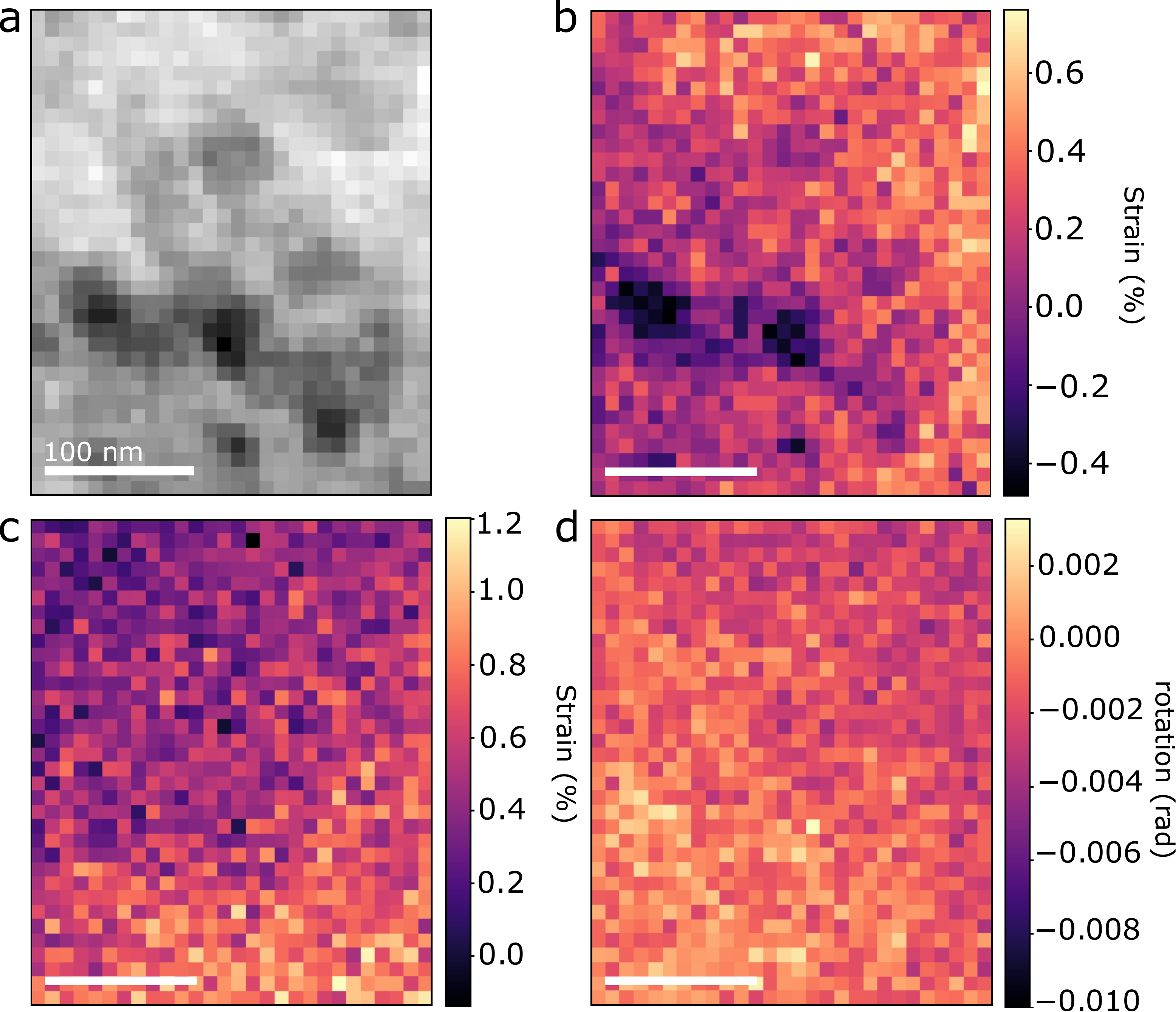}
		\caption{\textbf{Strain from same area as Fig. \ref{Trion_impurities}:} \textbf{(a)} MADF image, \textbf{(b)}, $\epsilon_{xx}$ component of the strain tensor, \textbf{(c)} $\epsilon_{yy}$ component of the strain tensor, \textbf{(d)} rotation angle extracted from the rotation matrix with polar decomposition. $\epsilon_{xy}$ corresponding to shear strain is not displayed and shows similar features as the rotation.
		The strain measurements below 1$\%$ show artefacts from the change in illumination in the diffraction spots. These maps illustrate that the diffraction measurements show only small variations that do not explain the trion localization by itself. }
		\label{SI_strain_hole_local}
	\end{figure}
	
	\begin{figure}
		\includegraphics[width=16cm]{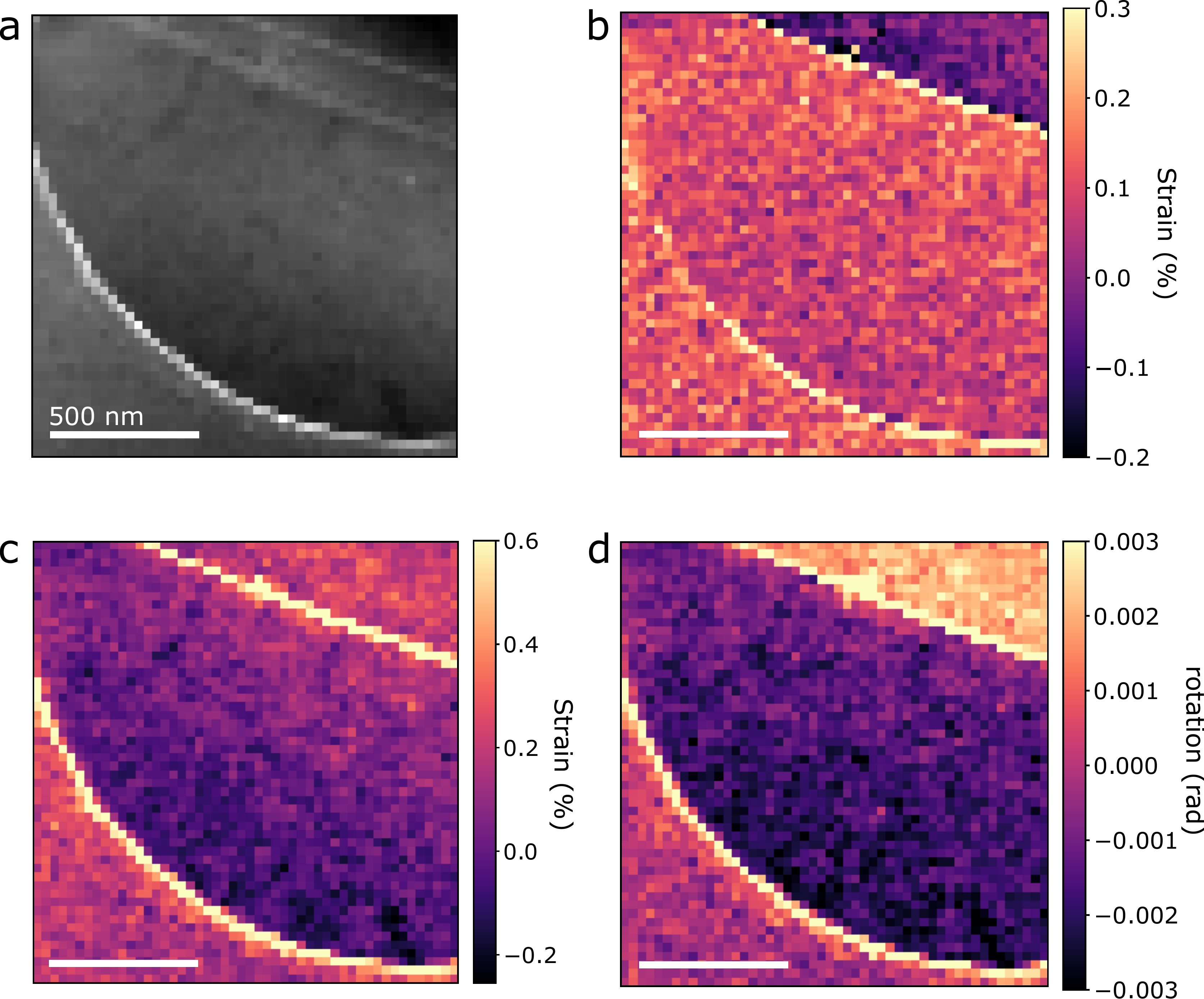}
		\caption{\textbf{Strain close to the carbon support in Fig. \ref{Trion_interface}:} \textbf{(a)} MADF image, \textbf{(b)}, $\epsilon_{xx}$ component of the strain tensor, \textbf{(c)} $\epsilon_{yy}$ component of the strain tensor, \textbf{(d)} rotation angle extracted from the rotation matrix with polar decomposition. $\epsilon_{xy}$ corresponding to shear strain is not displayed and shows similar features as the rotation.
		The strain measurements below 1$\%$ show artefacts from the change in illumination in the diffraction spots. These maps illustrate that the diffraction measurements show only small variations that do not explain the trion localisation by itself. }
		\label{SI_strain_hole_H1S1}
	\end{figure}

	\begin{figure}
		\includegraphics[width=16cm]{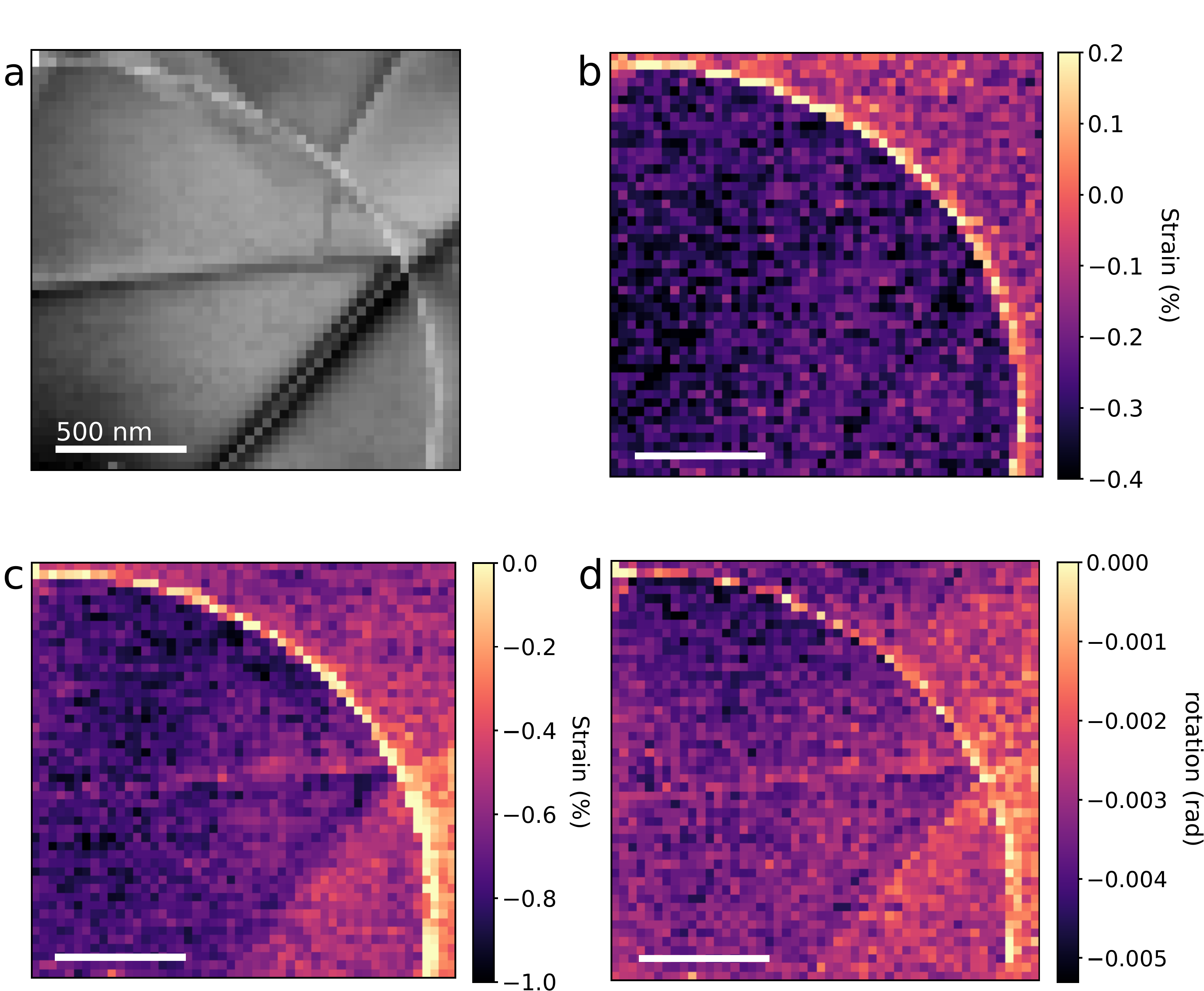}
		\caption{\textbf{Strain close to the carbon support in Fig. \ref{Local_Trion_L1}b:} \textbf{(a)} MADF image, \textbf{(b)}, $\epsilon_{xx}$ component of the strain tensor, \textbf{(c)} $\epsilon_{yy}$ component of the strain tensor, \textbf{(d)} rotation angle extracted from the rotation matrix with polar decomposition. $\epsilon_{xy}$ corresponding to shear strain is not displayed and shows similar features as the rotation.
		The strain measurements below 1$\%$ show artifacts from the change in illumination in the diffraction spots. These maps illustrate that the diffraction measurements show only small variations that do not explain the trion localisation by itself. }
		\label{SI_strain_hole_H3S1}
	\end{figure}
	
	\begin{figure}
		\includegraphics[width=16cm]{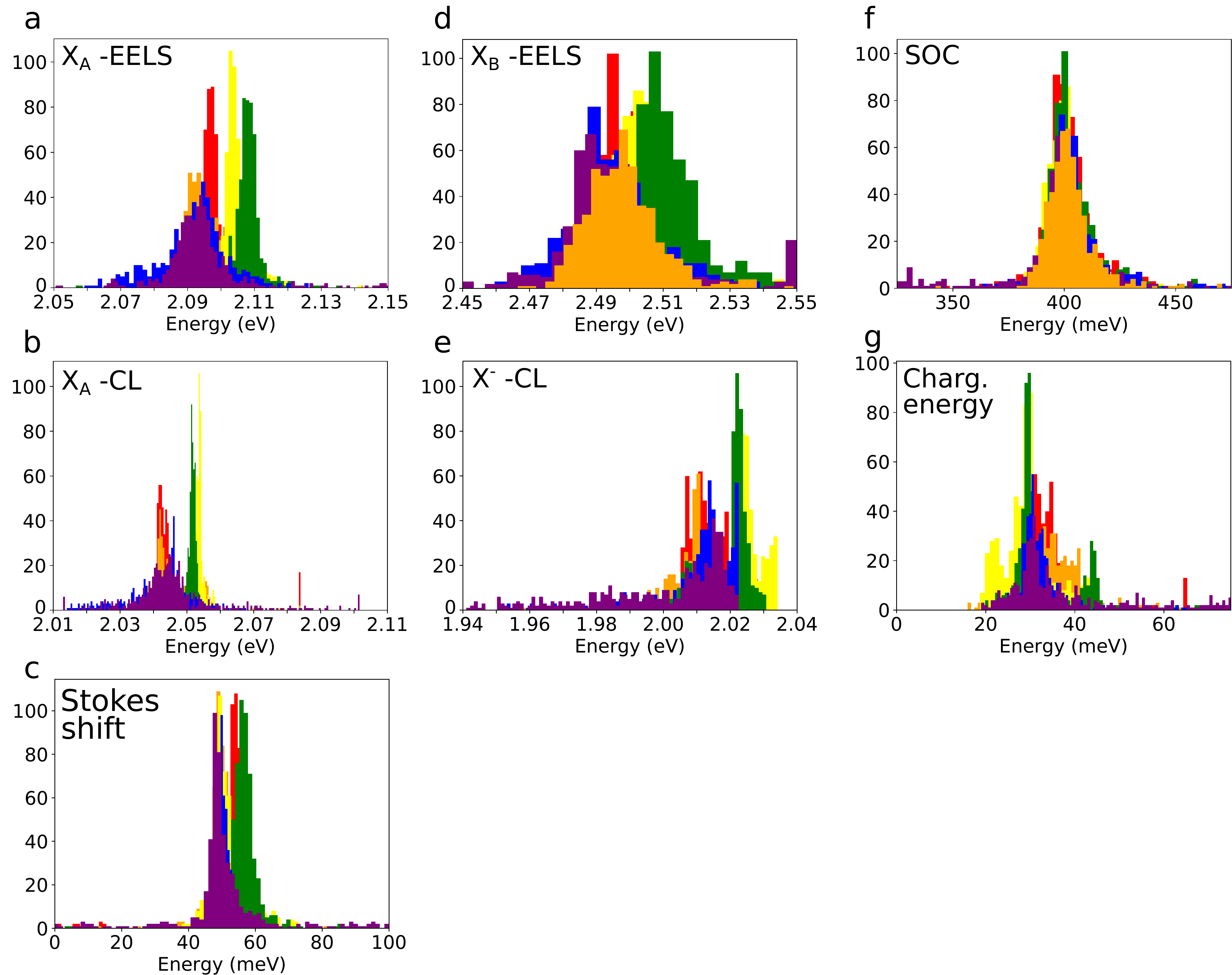}
		\caption{\textbf{Histogram of measured energies and energies differences in different regions:} \textbf{(a)} X$_A$ energy measured in EELS, \textbf{(b)} X$_A$ energy measured in CL, \textbf{(c)} X$_B$ energy measured in EELS, \textbf{(d)} X$^-$ energy measured in CL, \textbf{(e)} Stokes shift (corresponding to X$_{A,EELS}$ - X$_{A,CL}$ energy), \textbf{(f)} Spin-orbit splitting (corresponding to $X_B$ - X$_A$ energy), \textbf{(g)} charging energy (corresponding to X$_A$ - X$^-$ energy). The energies are extracted from Gaussian fit of each pixel of each datacube. The 12 datacubes were measured on the same day, on different areas of the same sample. The areas are of few µm$^2$ each. The colorcode refers to the different areas measured on the same heterostructure within holes in the amorphous carbon film.} 
		\label{SI_histograms_holes}
	\end{figure}
	
	\begin{figure}
		\includegraphics[width=12cm]{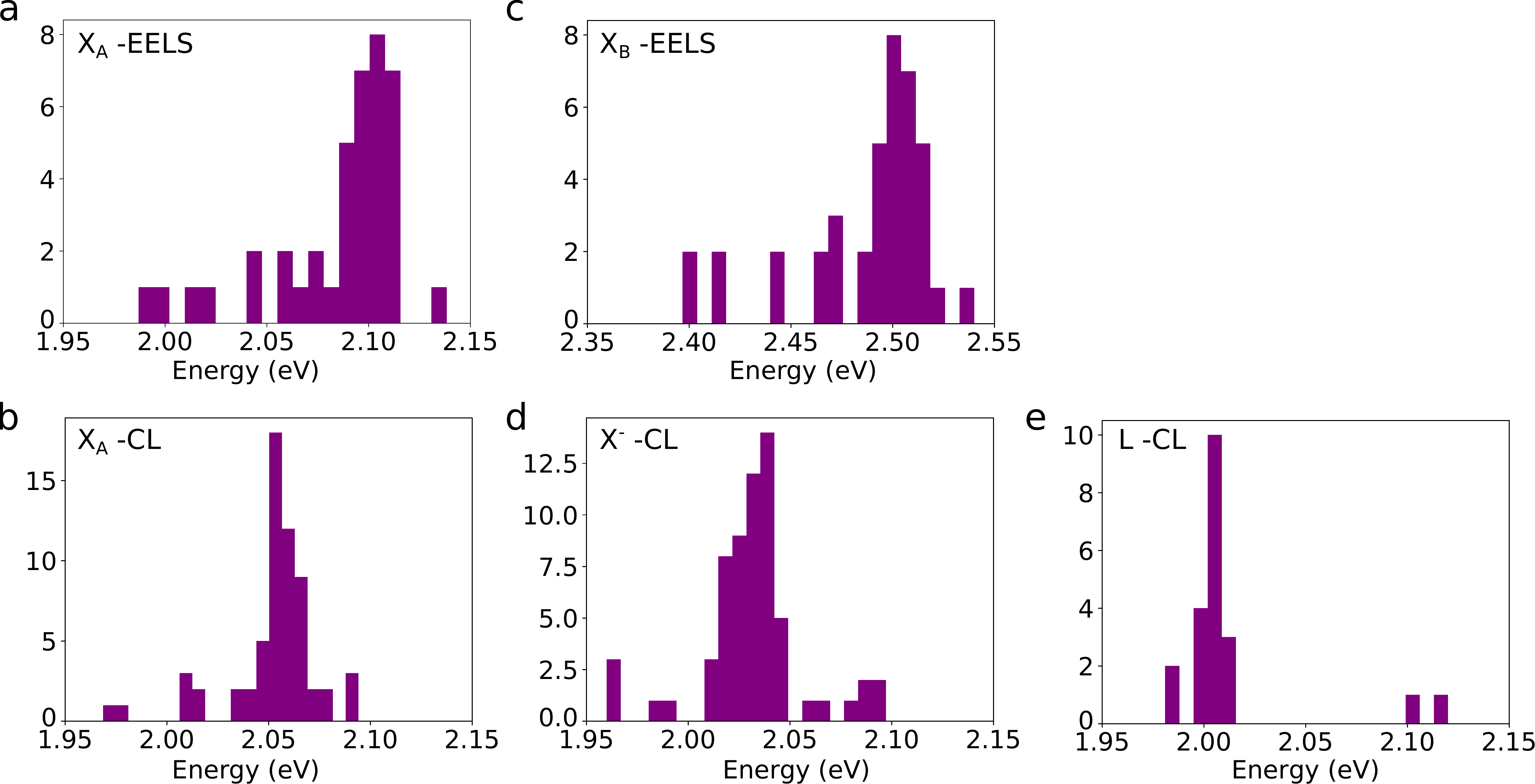}
		\caption{\textbf{Histogram of absorption and emission energies across all samples:} Each value of the histogram is the mean value of a measured datacube, each containing tens to hundreds of pixels. For EELS 40 datacubes are taken into account, for X$_A$-CL and X$^-$-CL 62 and for L-CL 21. The area covered by each datacube vary from tens of nanometers to few µm$^2$.   \textbf{(a)} X$_A$ energy measured in EELS, \textbf{(b)} X$_A$  energy measured in CL, \textbf{(c)} X$_B$ energy measured in EELS, \textbf{(d)} X$^-$ energy measured in CL. \textbf{(e)} L energy measured in CL. The dataset from which L is extracted is different than that of (b) and (c), to measure the L emission specifically.} 
		\label{SI_histograms}
	\end{figure}

\end{document}